\documentclass[aps,prx,showpacs,twocolumn,amsmath,amssymb]{revtex4-1}
\usepackage{graphicx}
\usepackage{dcolumn}
\usepackage{bm}
\usepackage{amsmath}
\usepackage{amsfonts}
\usepackage{amssymb}
\usepackage{multirow}

\usepackage{amsthm}
\usepackage{pinlabel}
\usepackage{natbib}

\usepackage{epstopdf}
\usepackage[abs]{overpic}
\usepackage[dvipsnames]{xcolor}

\usepackage[dvipsnames]{xcolor}

\newcommand{\blue}{\color{black}}

\newcommand{\no}{\nonumber}
\newcommand{\ph}{^{\phantom{\dagger}}}
\usepackage{braket}

\usepackage[normalem]{ulem}

\definecolor{specialgray}{HTML}{505050}
\definecolor{col10K}{HTML}{FFA000}
\definecolor{col300K}{HTML}{924FA4}
\definecolor{colMu}{HTML}{5278BD}
\definecolor{colMuI}{HTML}{924FA4}

\definecolor{specialgray}{HTML}{505050}
\definecolor{col10K}{HTML}{FFA000}
\definecolor{col300K}{HTML}{924FA4}
\definecolor{colMu}{HTML}{5278BD}
\definecolor{colMuI}{HTML}{924FA4}
\definecolor{newred}{HTML}{D53E4F}
\definecolor{newblue}{HTML}{5278BD}
\definecolor{newcyan}{HTML}{1EA0A0}
\definecolor{newgreen}{HTML}{5CB14E}
\definecolor{newpurple}{HTML}{924FA4}
\definecolor{newyellow}{HTML}{D1C72E}
\definecolor{neworange}{HTML}{D6923C}

\begin{document}

\title{Full-bandwidth Eliashberg theory of superconductivity beyond Migdal's approximation}
\author{Fabian Schrodi}\email{fabian.schrodi@physics.uu.se}
\author{Peter M. Oppeneer}\email{peter.oppeneer@physics.uu.se}
\author{Alex Aperis}\email{alex.aperis@physics.uu.se}
\affiliation{Department of Physics and Astronomy, Uppsala University, P.\ O.\ Box 516, SE-75120 Uppsala, Sweden}

\vskip 0.4cm
\date{\today}

\begin{abstract}
	\noindent 
 We solve the anisotropic, full-bandwidth and non-adiabatic Eliashberg equations for phonon-mediated superconductivity by fully including the first vertex correction in the electronic self-energy. The non-adiabatic equations are solved numerically here without further approximations, for a one-band model system. We compare the results to those that we obtain by adiabatic full-bandwidth, as well as Fermi-surface restricted Eliashberg-theory calculations. We find that non-adiabatic contributions to the superconducting gap can be positive, negative or negligible, depending on the dimensionality of the considered system, the degree of non-adiabaticity, and the coupling strength. We further examine non-adiabatic effects on the transition temperature and the electron-phonon coupling constant.  Our treatment emphasizes the importance of overcoming previously employed approximations in estimating the impact of vertex corrections on superconductivity and opens a pathway to systematically study vertex correction effects in systems such as high-$T_c$, flat band and low-carrier density superconductors.
\end{abstract}

\maketitle

\newpage

\section{Introduction}

The foundation for establishing the microscopic description of phonon mediated superconductors was laid by the  pioneering work of Migdal on the electron-phonon interaction in metals, where his famous theorem was introduced\,\cite{Migdal1958}.
The essence of Migdal's theorem lies in the effective {electron-phonon} coupling $\lambda$ times $\Omega/\epsilon_F$ being small, where $\Omega$ is the {characteristic} phonon frequency and $\epsilon_F$ the Fermi energy. Under this assumption it is possible to treat the {arising} infinite Feynman series of vertex diagrams in a perturbative manner with an expansion parameter $\lambda\Omega/\epsilon_F$. Migdal's theorem dictates that when $\Omega \ll \epsilon_F$ (or in other words, in the adiabatic limit), vertex corrections become negligible and the series can be truncated up to first-order in the coupling.
{For typical metals, where the degree of non-adiabaticity $\alpha \equiv \Omega/\epsilon_F\sim 10^{-2}$,} such an approximation to the electron self-energy (below referred to as Migdal's  approximation), is commonly believed to be valid even in materials characterized by strong-coupling $\lambda\gtrsim\mathcal{O}(1)$. 
 Eliashberg generalized this formalism to the superconducting state\,\cite{Eliashberg1960} and thereby laid the foundation for the thus-far most successful description of a vast amount of {superconductors that deviate from the weak-coupling limit of the Bardeen-Cooper-Schrieffer (BCS) theory \cite{Scalapino1966,Allen1983,carbotte1990,bennemann2008}.
}

However, the discovery of high-$T_c$ superconductors, for example the cuprates (for recent reviews, see, e.g., \cite{Damascelli2003,Keimer2015}), or monolayer FeSe on SrTiO$_3$ (FeSe/STO)\,\cite{Qing-Yan2012,Lee2014,Zhang2017}, has put the applicability of these conventional theories under debate \cite{Pietronero1992_1,Pietronero1995,Sadovskii2019}. This is mainly due to their common ingredients, such as a shallow electronic band near the Fermi level and comparatively large {boson  frequencies\,\cite{Lanzara2001,He2001,Liu2015,Wang2016c}}. 
{Other striking examples of materials that lie in the non-adiabatic regime include the record {h}igh-$T_c$ superconductor H$_3$S \cite{Drozdov2015,Errea2015,Jarlborg2016}, flat band systems like magic-angle twisted bilayer graphene \cite{cao2018}, and low-carrier density superconductors, like doped SrTiO$_3$ \cite{Marel2011,Lin2013}.}

The importance of vertex corrections has been studied theoretically only by a few groups\, \cite{Kostur1993,Gunnarsson1994,Nicol1994,Pietronero1995a,Grimaldi1995a,Miller1998,Alexandrov2001,Hague2003}, while numerical analyses were so far carried out mainly by Pietronero and coworkers\,\cite{Pietronero1992_2,Benedetti1994,Paci2001,Boeri2003,Cappelluti2003}. Recently, quantum Monte Carlo simulations have been employed to address the issue \cite{Esterlis2018}. Due to the huge numerical complexity of including vertex corrections in the electronic self-energy, studies so far have been subject to further approximations, 
{\blue such as taking the non-interacting limit of the vertex correction and averaging over momenta and therefore neglecting momentum dependence in the Eliashberg calculations \cite{Pietronero1995a,Grimaldi1995a}. {\blue Most o}f all, the 
standard practice is to integrate out and therefore neglect the contributions of electrons not at the vicinity of the Fermi surface.} 
Here we shall refer to {\blue the latter} type of calculations as Fermi surface restricted {(FSR)}. This includes the original works by Migdal and Eliashberg, too, and it is the reason why existing theories are often formulated in terms of the coupling $\lambda$, instead of the actual scattering strength $g_0$ arising in the Feynman expansion (see Section \ref{scMicroscopic} for details).

{\blue 
Migdal's theorem was originally formulated for 3D systems \cite{Migdal1958} and, based on phase-space arguments, it is generally expected to be violated for lower spatial dimensions \cite{Allen1983}. We will discuss this point in more detail further below. The available studies of low dimensionality effects on the vertex corrections are scarce and they 
{\blue notably} all make use of 
the aforementioned Fermi surface and momentum isotropy assumptions in order to tackle the problem analytically as in the original Migdal formulation. For example, the first vertex correction in the case of the 2D electron gas was found to be of order $\sim (\Omega/\epsilon_F)^{1/2}$ \cite{Madhukar1977} instead of the 3D result of $\sim (\Omega/\epsilon_F)$ whereas for 1D systems conclusions could not be drawn due to the enhanced tendency for lattice instabilities except from the special case of small-$q$ phonons where Migdal's theorem was shown to hold {\blue \cite{Apostol1982}}. 
Interestingly, the theorem breaks down in the case of such small-$q$ phonons in 3D systems, as Migdal himself pointed out \cite{Migdal1958,Allen1983}. 
Thus far, due to the lack of direct numerical calculations little is known about the effect of the momentum dependence in the Fermi surface shape and the electron energies on the solutions to the vertex-corrected Eliashberg equations.
Concomitantly, a systematic study of the effect of reduced spatial dimensionality on Migdal's approximation is still missing.} 

Our current work is motivated by the observation that many superconductors, including high-$T_c$'s, depart from the perfect adiabatic regime, hence there is no consensus about the smallness of vertex corrections to the bare electron-phonon scattering diagram. {\blue In addition, many of these materials are not purely 3D systems.}
Here, we address the question about the validity of the adiabatic approximation, using a less approximative treatment compared to available literature. It is worthwhile to establish under which conditions an adiabatic full bandwidth (AFB) \cite{Aperis2018,Schrodi2018,Gastiasoro2019} approach gives sufficiently accurate results, when even FSR  {Migdal-Eliashberg calculations \cite{Allen1983}} can be used for a successful description, and when the usage of non-adiabatic Eliashberg theory is inevitable. {\blue We also examine the possibility of applying a FSR isotropic approximation to the non-adiabatic Eliashberg equations {a}s an attempt to considerably decrease the high computational complexity of the problem. However, as we show, such simplifications often result in severe disagreement with the full non-adiabatic solutions and they are therefore generally not {acceptable.} Lastly, we provide a brief account on previous results and compare them with our direct numerical solutions.}

We assume conventional (s-wave) Cooper pairing and consider the first two {\blue infinite series of} Feynman diagrams for the electronic self-energy, from which it is possible to derive a set of self-consistent equations for the mass enhancement, the chemical potential renormalization, and the superconducting order parameter (given below in  Section \ref{scMicroscopic}). We numerically solve this system of equations without further approximation, i.e.\ the full momentum and frequency dependence is kept. In Section \ref{scSimp} we introduce the {AFB}, as well as the {FSR} equations which we solve to have reference points for comparison. {\blue For completeness, we also derive and solve the FSR isotropic Eliashberg equations with vertex corrections.} We use an effective one-band tight-binding model up to next-nearest neighbor hopping for the electronic energies. Our final model framework spans a huge parameter space, which allows systematic variations in the dimensionality, the characteristic phonon frequency, the strength of the coupling, the electronic bandwidth and, most importantly the degree of non-adiabaticity. {\blue In Section \ref{scDiagrams} we provide a heuristic discussion on the different approaches presented here on the basis of Feynman diagram expansions.} In Section \ref{scDimension} we study the influence of vertex corrections on the superconducting gap in systems of different dimensionality. We subsequently confine ourselves to 2D systems for which we examine similar parameter space with focus on the superconducting transition temperature and the electron-phonon coupling constant in Sec.\,\ref{scTrends}. {\blue We close {the}  Results Section with a discussion on previous approximations of the vertex function and {provide} a comparison with our results in Sec.\,\ref{scComparison}.} Our final conclusions and a brief outlook are given in Sec.\,\ref{scConclusion}.

\section{Theory}

We want to describe phonon-mediated superconductivity with a non-adiabatic and full bandwidth Eliashberg theory that goes beyond the commonly employed Migdal's approximation. As a starting point, we consider an electronic dispersion $\xi_{\mathbf{k}}^b$, rigidly shifted by a chemical potential $\mu$, such that $\xi_{\mathbf{k}}=\xi_{\mathbf{k}}^b-\mu${\blue , with $\mathbf{k}$ a  Brillouin zone (BZ) wave vector. For simplicity we focus on an effective one-band situation. As usual, {$\epsilon_F=|\underset{\mathbf{k}}{\mathrm{min}}\,\xi_{\mathbf{k}}|$} and  the `shallowness', i.e.\ the depth of the band, can be controlled by $\mu$ alone.}
Turning to the lattice vibrations, we assume an isotropic Einstein phonon spectrum with a {characteristic} frequency $\Omega$, see Sec.\,\ref{scMicroscopic} for details. By defining $\alpha=\Omega/\epsilon_F$ we can {associate} the electronic and bosonic energy scales {with} a parameter {$\alpha$} which is representative for the degree of non-adiabaticity in our system. 

\subsection{Microscopic description}\label{scMicroscopic}

In the following we use $b_{\mathbf{q}}$ and $b_{\mathbf{q}}^{\dagger}$ as bosonic annihilation and creation operator. The fermionic analogues $c_{\mathbf{k},\sigma}$ and $c^{\dagger}_{\mathbf{k},\sigma}$, with spin index $\sigma$, are hidden in Nambu spinors $\Psi^{\dagger}_{\mathbf{k}}=\big(c^{\dagger}_{\mathbf{k},\uparrow},c_{-\mathbf{k},\downarrow}\big)$ in the electron-phonon coupled Hamiltonian that reads
\begin{align}
H =& \sum_{\mathbf{k}}\xi_{\mathbf{k}}\Psi^{\dagger}_{\mathbf{k}} \hat{\rho}_3\Psi_{\mathbf{k}} + \sum_{\mathbf{q}}\hbar\Omega \left(b^{\dagger}_{\mathbf{q}}b_{\mathbf{q}} + \frac{1}{2}\right) \nonumber \\
&+\sum_{\mathbf{k},\mathbf{q}}g_{\mathbf{q}}u_{\mathbf{q}}\Psi^{\dagger}_{{\blue \mathbf{k-q}}}\hat{\rho}_3\Psi_{\mathbf{k}} ~. \label{hamiltonian}
\end{align}
In the above we include a purely electronic part, a bosonic term and a coupling between both. {Our focus here is solely on the effect of the latter term on superconductivity, therefore for the sake of simplicity we will not include the impact of Coulomb repulsion on the pairing.} Here we use the Pauli basis $\hat{\rho}_i$, $i=0,1,2,3$, and describe the ion displacement by {\blue $u_{\mathbf{q}}=(b^\dagger_{\bf q}+b\ph_{-{\bf q}})$}, while $g_{\mathbf{q}}$ {is the electron-phonon scattering strength.}
 
\begin{figure}[th!]
	\centering
	\begin{overpic}[width = 0.95\columnwidth, clip, unit=1pt]{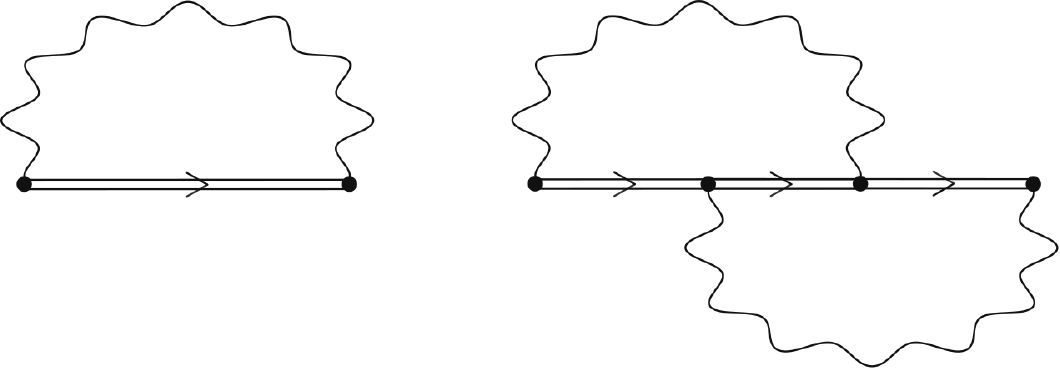}
		\put(-5, 75){{\large (a)}}
		\put(105, 75){{\large (b)}}
	\end{overpic}
	\caption{First ({a}) and second (b) order Feynman diagrams of the electron {\blue self-energy due to the electron-phonon interaction corresponding to the first and second term of Eq.\,(\ref{selfen}). Straight double lines are full electron Green's functions and wavy lines are bare phonon propagators.}}
	\label{feynman}
\end{figure}

 The electronic propagator as a function of imaginary time $\tau$ reads
\begin{align}
\hat{G}_{\mathbf{k}}(\tau) = - \langle \mathcal{T}_{\tau}\Psi_{\mathbf{k}}(\tau)\otimes  \Psi_{\mathbf{k}}^{\dagger}(0)\rangle ~,
\end{align}
where $\mathcal{T}_{\tau}$ is the time ordering operator. Let $\omega_m=\pi T(2m+1)$ be a fermionic Matsubara frequency with temperature $T$ and $m\in\mathbb{Z}$. The electron Green's function in Matsubara space obeys the Dyson equation
\begin{align}
\hat{G}_{\mathbf{k},m} = \hat{G}^0_{\mathbf{k},m} + \hat{G}^0_{\mathbf{k},m}\hat{\Sigma}_{\mathbf{k},m}\hat{G}_{\mathbf{k},m} ~,\label{dyson}
\end{align}
with the shorthand notation $F_{\mathbf{k},m} = F(\mathbf{k},i\omega_m)$ for any function $F$. The non-interacting Green's function is given by $[\hat{G}^0_{{\bf k},m}]^{-1}=i\omega_m\hat{\rho}_0-\xi_{{\bf k}}\hat{\rho}_3$. For evaluating the electronic self-energy $\hat{\Sigma}_{\mathbf{k},m}$ we consider in Fig.\,\ref{feynman} the first and second order scattering diagrams which are shown in panels (a) and (b), respectively.
{\blue A straight arrowed double line represents the renormalized electron Green's function of Eq.\ (\ref{dyson}) and each wavy line the phonon propagator.} Each vertex pair is associated with a factor $|g_{\mathbf{q}}|^2\equiv |g_0f_{\mathbf{q}}|^2$, which we rewrite as product of scattering strength $g_0$ and form factor $f_{\mathbf{q}}$ {($\rm{max}\,|f_{\mathbf{q}}|=1$). The latter carries the information of the momentum dependence of the interaction}. The expansion parameter in the Feynman diagram series is therefore $|g_0|^2$. 

Taking into account both diagrams of Fig.\,\ref{feynman} 
we find the anisotropic electronic self-energy as

\begin{align}
\hat{\Sigma}_{{\bf k}m}&= T\sum_{{\bf k}'m'}D_{{\bf k}-{\bf k}',m-m'}|g_{{\bf k}-{\bf k}'}|^2\hat{\rho}_3 \hat{G}_{{\bf k}',m'}\hat{\rho}_3 \nonumber \\
& +   T^2 \sum_{{\bf k}'m'}\sum_{{\bf k}''m''}D_{{\bf k}-{\bf k}',m-m'}D_{{\bf k}'-{\bf k}'',m'-m''} \nonumber \\
&  \times |g_{{\bf k}-{\bf k}'}|^2|g_{{\bf k}'-{\bf k}''}|^2\hat{\rho}_3\hat{G}_{{\bf k}',m'}\hat{\rho}_3\hat{G}_{{\bf k}'',m''} \nonumber \\
& \times \hat{\rho}_3\hat{G}_{{\bf k}''-{\bf k}'+{\bf k},m''-m'+m}\hat{\rho}_3 ~. \label{selfen} 
\end{align} 
{\blue The first term on the right-hand-side of Eq.\ (\ref{selfen}) corresponds to the diagram of Fig.\ \ref{feynman}(a) and describes an infinite series of non-crossing, so-called rainbow diagrams. Keeping only this term in the electronic self-energy is known as Migdal's approximation. The second term on the right-hand-side of Eq.\ (\ref{selfen}) corresponds to the diagram of Fig.\ \ref{feynman}(b) and describes an infinite series of crossed  diagrams. Inclusion of this term yields the first vertex correction to the electronic self-energy beyond Migdal's approximation.}
Due to momentum conservation we use $\mathbf{q}=\mathbf{k}-\mathbf{k}'$ in Eq.\,(\ref{selfen}). As indicated before, the scattering amplitudes $|g_{{\bf k}-{\bf k}'}|^2$ and phonon propagators $D_{{\bf k}-{\bf k}',m-m'}$ are assumed to be branch and band independent. {The Einstein-like phonon spectrum that we consider here leads to}
%
\begin{eqnarray}\no
D_{{\bf k}-{\bf k}',m-m'}|g_{{\bf k}-{\bf k}'}|^2 &=& \!\!\int_{0}^{\infty}\! \frac{{d}\omega}{N_0}\alpha^2F({\bf k},{\bf k}',\omega)\\\no
&\times&\frac{2\omega}{(\omega_m-\omega_{m'})^2 + \omega^2} \\
&\equiv& V_{{\bf k}-{\bf k}',m-m'} ~, \label{int}
\end{eqnarray}
with $V_{{\bf k}-{\bf k}',m-m'}$ the electron-phonon interaction kernel. In Eq.\,(\ref{int}) the electron density of states {\blue (DOS)} at the Fermi level is given by $N_0$, and $\alpha^2F({\bf k},{\bf k}')=N_0|g_{{\bf k}-{\bf k}'}|^2\delta(\omega-\Omega)$ is the momentum resolved Eliashberg function.

From here we follow the standard recipe of Eliashberg theory and introduce the mass enhancement $Z_{{\bf k},m}$, the chemical potential renormalization $\chi_{{\bf k},m}$, and the superconducting order parameter $\phi_{{\bf k},m}$ by writing
\begin{align}
\hat{G}_{\mathbf{k},m}^{-1} \equiv i\omega_mZ_{\mathbf{k},m}\hat{\rho}_0 - \big(\xi_{\mathbf{k}}+\chi_{\mathbf{k},m}\big)\hat{\rho}_3 - \phi_{\mathbf{k},m}\hat{\rho}_1~. \label{defGreen}
\end{align}
By combining Eqs.\,(\ref{dyson}-\ref{defGreen}) it is possible to project on channels $\hat{\rho}_0$, $\hat{\rho}_3$ and $\hat{\rho}_1$, and find the self-consistent equations for corresponding prefactors in Eq.\,(\ref{defGreen}). After some tedious algebra we obtain
\begin{widetext}
	\begin{align}
	Z_{{\bf k},m} &= 1 - \frac{T}{\omega_m} \sum_{{\bf k}'m'} V_{{\bf k}-{\bf k}',m-m'} \left(\gamma_{{\bf k}',m'}^{(Z)} + T\sum_{{\bf k}''m''} V_{{\bf k}'-{\bf k}'',m'-m''} \vec{\gamma}_{{\bf k}'',m''}^T P_{{\bf k}',m'}^{(Z)} \vec{\gamma}_{{\bf k}''-{\bf k}'+{\bf k},m''-m'+m} \right) , \label{z}   \\
	\chi_{{\bf k},m} &= T \sum_{{\bf k}'m'} V_{{\bf k}-{\bf k}',m-m'}\left(\gamma_{{\bf k}',m'}^{(\chi)}  + T\sum_{{\bf k}''m''} V_{{\bf k}'-{\bf k}'',m'-m''} \vec{\gamma}_{{\bf k}'',m''}^T P_{{\bf k}',m'}^{(\chi)} \vec{\gamma}_{{\bf k}''-{\bf k}'+{\bf k},m''-m'+m} \right)  ,  \label{chi} \\
	\phi_{{\bf k},m} &= -T \sum_{{\bf k}'m'} V_{{\bf k}-{\bf k}',m-m'}\left(\gamma_{{\bf k}',m'}^{(\phi)}  + T\sum_{{\bf k}''m''} V_{{\bf k}'-{\bf k}'',m'-m''} \vec{\gamma}_{{\bf k}'',m''}^T P_{{\bf k}',m'}^{(\phi)} \vec{\gamma}_{{\bf k}''-{\bf k}'+{\bf k},m''-m'+m} \right) ~. \label{phi}
	\end{align}
%
\end{widetext}
For brevity we define the pseudo vector $\vec{\gamma}_{{\bf k},m}^T=(\gamma_{{\bf k},m}^{(Z)},\gamma_{{\bf k},m}^{(\chi)},\gamma_{{\bf k},m}^{(\phi)})$ with elements $\gamma_{{\bf k},m}^{(Z)}=\omega_mZ_{{\bf k},m}/\Theta_{{\bf k},m}$, $\gamma_{{\bf k},m}^{(\chi)}=(\chi_{{\bf k},m}+\xi_{{\bf k}})/\Theta_{{\bf k},m}$ and $\gamma_{{\bf k},m}^{(\phi)}=\phi_{{\bf k},m}/\Theta_{{\bf k},m}$. The denominators are given by
\begin{align}
\Theta_{{\bf k},m} = (i\omega_m)^2Z_{{\bf k},m}^2-(\chi_{{\bf k},m}+\xi_{{\bf k}})^2-\phi_{{\bf k},m}^2 ~,
\end{align}
while the matrices in Eqs.\,(\ref{z}-\ref{phi}) have the form
\renewcommand{\arraystretch}{1}
\begin{align}
\setlength\arraycolsep{1pt}
P_{{\bf k},m}^{(Z)} &=  \begin{pmatrix}
-\gamma_{{\bf k},m}^{(Z)} & \gamma_{{\bf k},m}^{(\chi)} & \gamma_{{\bf k},m}^{(\phi)} \\
\gamma_{{\bf k},m}^{(\chi)} & \gamma_{{\bf k},m}^{(Z)} & 0 \\
-\gamma_{{\bf k},m}^{(\phi)} & 0 & -\gamma_{{\bf k},m}^{(Z)}
\end{pmatrix}  ,~ \\ 
P_{{\bf k},m}^{(\chi)} &= \begin{pmatrix}
-\gamma_{{\bf k},m}^{(\chi)} & -\gamma_{{\bf k},m}^{(Z)} & 0 \\
-\gamma_{{\bf k},m}^{(Z)} & \gamma_{{\bf k},m}^{(\chi)} & -\gamma_{{\bf k},m}^{(\phi)} \\
0 & -\gamma_{{\bf k},m}^{(\phi)} & -\gamma_{{\bf k},m}^{(\chi)}
\end{pmatrix} ,~ 
\\ 
P_{{\bf k},m}^{(\phi)} &= \begin{pmatrix}
-\gamma_{{\bf k},m}^{(\phi)} & 0 & -\gamma_{{\bf k},m}^{(Z)} \\
\setlength\arraycolsep{1pt}
0 & \gamma_{{\bf k},m}^{(\phi)} & \gamma_{{\bf k},m}^{(\chi)} \\
\gamma_{{\bf k},m}^{(Z)} & \gamma_{{\bf k},m}^{(\chi)} & -\gamma_{{\bf k},m}^{(\phi)}
\end{pmatrix} .
\end{align}
The superconducting gap can be found from $\Delta_{\mathbf{k},m}=\phi_{\mathbf{k},m}/Z_{\mathbf{k},m}$. 
{The diagram in Fig.\,\ref{feynman}{(a)} corresponds to the left, while the diagram in Fig.\,\ref{feynman}{(b)} {corresponds} to the right summand in the brackets of Eqs.\,(\ref{z}-\ref{phi}).}

{Taking a step further from} existing theories we {purposely} did not perform any {further energy integrations} or other approximations in deriving Eqs.\,(\ref{z}-\ref{phi}), hence our treatment is formally exact up to second order in $|g_0|^2$ {\blue in the expansion shown in Fig.\,\ref{feynman}}. Keeping all dependencies as they are stated above we solve for $Z_{\mathbf{k},m}$, $\chi_{\mathbf{k},m}$ and $\phi_{\mathbf{k},m}$ in an iterative self-consistent loop; the results are presented in Sec.\,\ref{scResults} below
and we give further numerical details in Appendix \ref{appNum}. The algorithm for solving the non-adiabatic, anisotropic and full bandwidth Eliashberg equations has been included in the Uppsala superconductivity (UppSC) code\,\cite{Aperis2015,Aperis2018,Schrodi2018,Schrodi2019,Bekaert2018,UppSC}.

\subsection{Simplifications to the model}\label{scSimp}

Before going to our numerical results it is useful to briefly discuss some limiting cases, with which we want to compare. 

{\blue \textit{Adiabatic full-bandwidth (AFB)  equations}.} From Eqs.\,(\ref{z}-\ref{phi}) it is straightforward to obtain the adiabatic limit, in which we can neglect all contributions from the {\blue diagram} in Fig.\,\ref{feynman}{(b)}. This results in the well-known adiabatic Eliashberg equations
\begin{align}
Z_{{\bf k},m}^{(\mathrm{ad})} &= 1 - \frac{T}{\omega_m} \sum_{{\bf k}'m'} V_{{\bf k}-{\bf k}',m-m'} \frac{\omega_{m'}Z_{{\bf k}',m'}^{(\mathrm{ad})}}{\Theta_{{\bf k}',m'}^{(\mathrm{ad})}}  ,  \label{zAd}  \\
\chi_{{\bf k},m}^{(\mathrm{ad})} &= T \sum_{{\bf k}'m'} V_{{\bf k}-{\bf k}',m-m'}\frac{\chi_{{\bf k}',m'}^{(\mathrm{ad})}+\xi_{{\bf k}'}}{\Theta_{{\bf k}',m'}^{(\mathrm{ad})}}   , \label{chiAd} \\
\phi_{{\bf k},m}^{(\mathrm{ad})} &= -T \sum_{{\bf k}'m'} V_{{\bf k}-{\bf k}',m-m'}\frac{\phi_{{\bf k}',m'}^{(\mathrm{ad})}}{\Theta_{{\bf k}',m'}^{(\mathrm{ad})}}  ~, \label{phiAd}
\end{align}
where we use the label (ad) to avoid confusion with the non-adiabatic functions. Consequently, we obtain the gap function $\Delta_{\mathbf{k},m}^{(\mathrm{ad})}=\phi_{\mathbf{k},m}^{(\mathrm{ad})}/Z_{\mathbf{k},m}^{(\mathrm{ad})}$.

{\blue \textit{Fermi-surface restricted (FSR) adiabatic   equations}.} In cases where only processes at the Fermi level are relevant to describe the interacting system it is possible to derive a further simplified, but still anisotropic theory. 
Using label (Fs), for Fermi surface, the mass renormalization and gap function for these systems can be given as
\begin{align}
Z_{\mathbf{k},m}^{(\mathrm{Fs})} &= 1 + \frac{\pi T}{\omega_m} \sum_{\mathbf{k}'m'}\lambda_{\mathbf{k}-\mathbf{k}',m-m'} \frac{\delta(\xi_{\mathbf{k}'})}{N_0} \frac{\omega_{m'}}{\theta_{\mathbf{k}',m'}} ,\label{zFs} \\
\Delta_{\mathbf{k},m}^{(\mathrm{Fs})} &= \frac{\pi T}{Z_{\mathbf{k},m}^{(\mathrm{Fs})}} \sum_{\mathbf{k}'m'}\lambda_{\mathbf{k}-\mathbf{k}',m-m'} \frac{\delta(\xi_{\mathbf{k}'})}{N_0} \frac{\Delta_{\mathbf{k}',m'}^{(\mathrm{Fs})}}{\theta_{\mathbf{k}',m'}} ,\label{deltaFs}
\end{align}
with $\theta_{\mathbf{k},m}=\sqrt{\omega_m^2+\big(\Delta_{\mathbf{k},m}^{(\mathrm{Fs})}\big)^2}$ and the coupling $\lambda_{\mathbf{k}-\mathbf{k}',m-m'}=N_0V_{\mathbf{k}-\mathbf{k}',m-m'}$. In deriving Eqs.\,(\ref{zFs}) and (\ref{deltaFs}) one assumes an infinite electronic bandwidth of a system at half-filling \cite{Allen1983}.

{\blue \textit{Fermi-surface restricted, isotropic approximation to the non-adiabatic equations}.} 
{\blue For completeness, we will also study the case where we apply the same approximations as in the derivation of the FSR adiabatic equations to the non-adiabatic Eqs\,(\ref{z}-\ref{phi}). Given the very high computational complexity of numerically solving Eqs.\,(\ref{z}-\ref{phi}), it is worth investigating how well such simplified non-adiabatic equations compare with the full Eqs.\,(\ref{z}-\ref{phi}). For the isotropic electron-phonon interaction that we adopt here, the equations derived below can be considered as the non-adiabatic extensions of Eqs.\,(\ref{zFs}-\ref{deltaFs}). As we show in Sec.\,\ref{scResults}, such simplified  non-adiabatic equations are unfortunately not reliable.

We focus on the second term on the right-hand side of Eq.\,(\ref{selfen}). Following the same procedure as in deriving Eqs.\,(\ref{z}-\ref{phi}) we obtain the following expression for the non-adiabatic part of the self-energy:
\begin{eqnarray}\no
\hat{\Sigma}_{m}^{(\mathrm{iso)}} &=& T^2 P \sum_{m',m''} \lambda_{m-m'} \lambda_{m'-m''}\\
&\times &\hat{\rho}_3 \hat{g}_{m'} \hat{\rho}_3 \hat{g}_{m''} \hat{\rho}_3 \hat{g}_{m''-m'+m} \hat{\rho}_3 ~,
\end{eqnarray}
where we use the label $(\mathrm{iso})$ for `isotropic'. The constant $P$ is defined as
\begin{align}\label{momcons}
P = \frac{1}{N_0^3} \sum_{\mathbf{k},\mathbf{k}',\mathbf{k}''} \delta(\xi_{\mathbf{k}})\delta(\xi_{\mathbf{k}'})\delta(\xi_{\mathbf{k}''})\delta(\xi_{\mathbf{k}''-\mathbf{k}'+\mathbf{k}}) ,
\end{align}
and reflects momentum conservation of processes at the Fermi level. The propagator $\hat{g}_m$ is obtained  by integrating the electron Green's function over energy,
\begin{align}
\hat{g}_{\mathbf{k},m} = \int_{-\infty}^{\infty} {d}\epsilon\,  \hat{G}_{\mathbf{k},m}(\epsilon)  ~,
\end{align} 
and the result is further{more} assumed to be isotropic, $\hat{g}_{\mathbf{k},m}\simeq\hat{g}_m$. We get
\begin{align}
\hat{g}_{m} = -\pi \frac{i\omega_m\hat{\rho}_0 + \Delta_m^{(\mathrm{iso})}\hat{\rho}_1}{\sqrt{\omega_m^2 + (\Delta_{m}^{(\mathrm{iso})})^2}} ~.
\end{align}
With the above considerations, the resulting Eliashberg equations are
\begin{eqnarray}\no
Z_m^{(\mathrm{iso})} &=& 1 + \frac{\pi T}{\omega_m} \sum_{m'} \lambda_{m-m'} \gamma_{m'}^{(\omega)} \\\label{iso1}
&+& P \frac{\pi^3T^2}{\omega_m} \sum_{m,m'} \lambda_{m-m'}\lambda_{m'-m''} \\\no
&\times&\Bigg( -\gamma_{m'}^{(\omega)}\gamma_{m''}^{(\Delta)}\gamma_{m''-m'+m}^{(\Delta)} 
+ \gamma_{m'}^{(\Delta)}\gamma_{m''}^{(\omega)}\gamma_{m''-m'+m}^{(\Delta)} \\\no
&-& \gamma_{m'}^{(\Delta)}\gamma_{m''}^{(\Delta)}\gamma_{m''-m'+m}^{(\omega)} - \gamma_{m'}^{(\omega)}\gamma_{m''}^{(\omega)}\gamma_{m''-m'+m}^{(\omega)} \Bigg), \\\no
\Delta_m^{(\mathrm{iso})} &=& \frac{\pi T}{Z_m^{(\mathrm{iso})}} \sum_{m'} \lambda_{m-m'} \gamma_{m'}^{(\Delta)} \\\label{iso2}
&+& P \frac{\pi^3T^2}{Z_m^{(\mathrm{iso})}} \sum_{m',m''} \lambda_{m-m'}\lambda_{m'-m''}\\\no
&\times& \Bigg( -\gamma_{m'}^{(\Delta)}\gamma_{m''}^{(\Delta)}\gamma_{m''-m'+m}^{(\Delta)} 
- \gamma_{m'}^{(\omega)}\gamma_{m''}^{(\omega)}\gamma_{m''-m'+m}^{(\Delta)} \\\no
&+& \gamma_{m'}^{(\omega)}\gamma_{m''}^{(\Delta)}\gamma_{m''-m'+m}^{(\omega)} - \gamma_{m'}^{(\Delta)}\gamma_{m''}^{(\omega)}\gamma_{m''-m'+m}^{(\omega)}\Bigg) ~.
\end{eqnarray}
Above we use $\gamma_{m}^{(\omega)}=\omega_m/\sqrt{\omega_m^2 + (\Delta_m^{(\mathrm{iso})})^2}$ and $\gamma_{m}^{(\Delta)}=\Delta^{(\mathrm{iso})}_m/\sqrt{\omega_m^2 + (\Delta_m^{(\mathrm{iso})})^2}$.
}

In Sec.\,\ref{scResults}, we perform a parameter space exploration following each of these {\blue four} different approaches. We have the {\blue {AFB} Eqs.\,(\ref{zAd}-\ref{phiAd}), the FSR Eqs.\,(\ref{zFs}-\ref{deltaFs}),  the full bandwidth non-adiabatic description, that includes both diagrams of Fig.\,\ref{feynman}, in Eqs.\,(\ref{z}-\ref{phi}) and lastly the isotropic FSR approximated non-adiabatic Eqs.\,(\ref{iso1}-\ref{iso2}). These last two sets of  equations are the respective non-adiabatic extensions including the first vertex correction of the former two sets of Migdal-Eliashberg equations.}

{\blue \textit{Details on the solved model.}} For the electronic energies we use a tight-binding description up to {the} next-nearest neighbor, reading
\begin{align}
\xi_{\mathbf{k}} =& -\sum_{i=x,y,z} t^{(1)}_i\cos(k_i) \nonumber \\
& - \sum_{i=x,y,z}\prod_{j=x,y,z; j\neq i} t_j^{(2)}\cos(k_j) - \mu ~. \label{dispInit}
\end{align}
For simplicity we assume equal hopping {energies} along all three spatial directions and write $t_i^{(1)}=t^{(1)}c_i$, $t_i^{(2)}=\sqrt{t^{(2)}} c_i$. The $c_i$ are conveniently used to control the dimensionality (from 1D to 3D) of our system, hence $c_x=1$ and $c_{i\neq x}\in[0,1]$. Inserting in Eq.\,(\ref{dispInit}) gives
\begin{align}
\xi_{\mathbf{k}} =& - t^{(1)}\sum_{i=x,y,z} c_i\cos(k_i) \nonumber \\
& - t^{(2)}\sum_{i=x,y,z}\prod_{j=x,y,z; j\neq i} c_j\cos(k_j) - \mu ~,
\end{align}
from which we find the electronic bandwidth as $W = \big| \underset{\mathbf{k}}{\mathrm{max}}\,\xi_{\mathbf{k}} - \underset{\mathbf{k}}{\mathrm{min}}\,\xi_{\mathbf{k}} \big|$. By fixing $W$ we are able to obtain the nearest-neighbor hopping from
\begin{align}
t^{(1)} = W \bigg/ \bigg(2\sum_{i=x,y,z}c_i\bigg) ~.
\end{align}
We further take $t^{(2)} = t^{(1)}/2$ as the next-nearest neighbor hopping. For an initial choice of characteristic phonon frequency $\Omega$ and degree of non-adiabaticity $\alpha$ one can deduce the shallowness of our dispersion $\xi_{\mathbf{k}}$ by {$\epsilon_F=\Omega/\alpha=|\underset{\mathbf{k}}{\mathrm{min}}\,\xi_{\mathbf{k}}|$}, which uniquely determines the global chemical potential $\mu$.

Another degree of freedom in our calculations is the coupling strength, which is commonly measured as $\lambda$, and treated here as parameter that is varied from weak to strong coupling. As stated in Sec.\,\ref{scMicroscopic}, the momentum dependence of the interaction kernel is $g_{\mathbf{q}}=g_0f_{\mathbf{q}}$, so we can write
\begin{align}
\lambda_{\mathbf{q},m-m'} = N_0V_{\mathbf{q},m-m'} = N_0 \frac{|g_0 f_{\mathbf{q}}|^2}{\Omega} \frac{ 2 \Omega^2}{\Omega^2+q^2_{m-m'}} ~. \label{lambdaq}
\end{align} 
In the case of a {FSR} calculation, as in Eqs.\,(\ref{zFs}) and (\ref{deltaFs}), the coupling can be extracted by a double momentum average at the Fermi level:
\begin{align}
\lambda = \langle\langle \lambda_{\mathbf{q},m-m'=0}\rangle_{\mathbf{k}_F^{ }}\rangle_{\mathbf{k}_F'} . \label{fsaverage}
\end{align}
Combining Eqs.\,(\ref{lambdaq}) and (\ref{fsaverage}) allows us to solve for the scattering strength
\begin{align}
|g_0|^2 = \frac{\lambda\Omega}{2N_0} \frac{1}{\langle\langle |f_{\mathbf{q}}|^2\rangle_{\mathbf{k}_F^{ }}\rangle_{\mathbf{k}_F'}} ~, \label{g0}
\end{align}
which can then be used to calculate a kernel corresponding to $\lambda$ for the adiabatic or non-adiabatic full-bandwidth equations.

To summarize, within the here presented setup we have the freedom to choose the systems dimensionality via $c_i$, the characteristic phonon frequency $\Omega$, the degree of non-adiabaticity $\alpha$, the coupling strength $\lambda$ and the momentum structure of the electron-phonon interaction. 
{All subsequent results follow from the considerations in this section.}

{\blue \subsection{Diagrammatic analysis}\label{scDiagrams}}

\begin{figure}[ht!]
	\centering
	\begin{overpic}[width = 1\columnwidth, clip, unit=1pt]{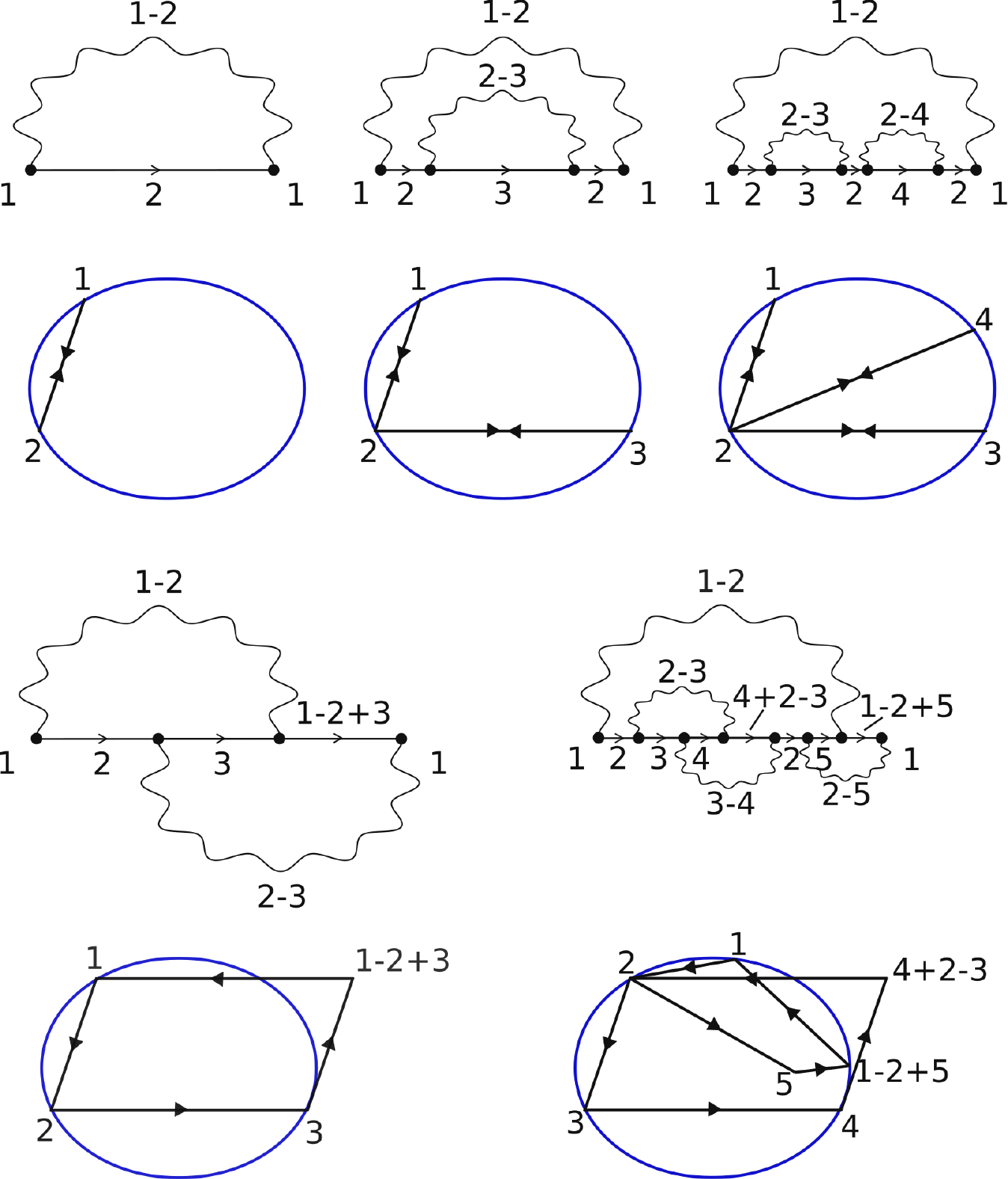}
		\put(0, 280){{\large (a)}}
		\put(80, 280){{\large (b)}}
		\put(165, 280){{\large (c)}}
		\put(0, 220){{\large (d)}}
		\put(80, 220){{\large (e)}}
		\put(165, 220){{\large (f)}}
		\put(0, 140){{\large (g)}}
		\put(130, 140){{\large (h)}}
		\put(0, 55){{\large (i)}}
		\put(130, 55){{\large (j)}}
	\end{overpic}
	\caption{Lowest order diagrams contributing to the electron self-energy of Fig.\,\ref{feynman}. {Graphs} (a), (b), and (c) are examples of  non-crossing diagrams retained in Migdal's approximation. {Plots} (d), (e), and (f) show Fermi surface sketches (blue ellipses) where arrows depict scattering processes corresponding to panels (a), (b), and (c), respectively. {Graphs} (g) and (h) are examples of the first vertex corrected crossing diagrams beyond Migdal's approximation. {Plots} (i) and (j) depict the corresponding scattering processes near the Fermi surface.}
	\label{feynman2}
\end{figure}
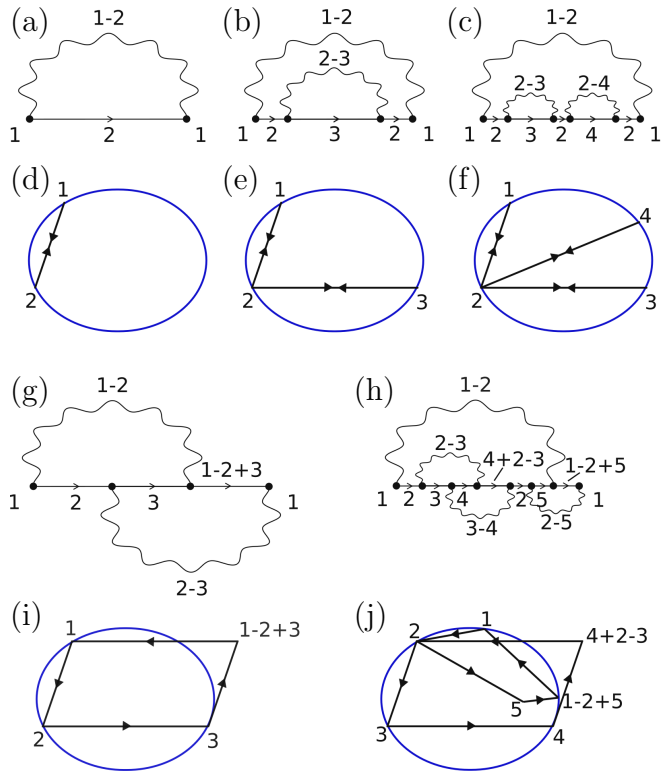

Before proceeding to present our numerical results in Sec.\,\ref{scResults}, we will elucidate further each of the approaches discussed in the previous section by means of a heuristic discussion based on Feynman diagrams. Fig.\,\ref{feynman2} shows a few low-order self-energy diagrams in terms of the non-interacting electron Green's function of Eq.\,(\ref{dyson}) that are contained in the infinite series implied in the diagrams of Fig.\,\ref{feynman}. These bare electron propagators are drawn with straight arrowed lines whereas the phonon propagators are drawn with wavy lines, as usual. Bare vertices are drawn with solid circles. The numbers in Fig.\,\ref{feynman2} designate four-momenta carried by respective propagators, i.e. $1\equiv({\bf k}_1,\omega_{m_1})$, $1-2+3\equiv({\bf k}_1-{\bf k}_2+{\bf k}_3,\omega_{m_1}-\omega_{m_2}+\omega_{m_3})$, etc. Although we do not draw the ingoing/outgoing electron propagators we nevertheless have chosen to show their corresponding four-momenta so as to keep track of the momentum and energy conservation for each diagram. 

The blue ellipses below each diagram are Fermi surface cartoons. {Straight lines with arrows} connecting two such four-momenta in these cartoons represent particle scattering events. These  arrow lines can be seen as vectors whose direction and amplitude follows from momentum and energy conservation at each vertex as we read the diagram from left to right. Lines connecting points on the ellipse describe scattering at the Fermi surface whereas for processes away from the Fermi level the edge of these lines moves further apart from the blue {ellipse}. Increasing the non-adiabatic ratio $\alpha$ in this picture simply means that these lines are allowed to extend further away from the Fermi surface {ellipse}, \textit{i.e.\ the phase-space for processes relevant to each diagram increases}. 

Following the discussion of Ref.\,\cite{Allen1983} for the non-superconducting state, and for the sake of simplicity, we neglect any contribution of the phonon propagator poles and consider that the major contributions in Eq.\,(\ref{selfen}) come from poles in the electron Green's functions. The closer to the Fermi level, the more pronounced the poles of these Green's functions are. Therefore, processes with lines connecting points on the ellipses are dominant. Diagrams with more lines, i.e.\ with more scattering events become less likely to contain sharp poles either because of eventually reduced phase-space or because they inherently involve processes away from the Fermi level.

Fig.\,\ref{feynman2}(a) shows the lowest-order non-crossing diagram of the adiabatic self-energy. Particles with $({\bf k}_1,\omega_{m_1})$ scatter off a virtual phonon to a state $({\bf k}_2,\omega_{m_2})$ before scattering back to their initial state $({\bf k}_1,\omega_{m_1})$. This is shown pictorially in Fig.\,\ref{feynman2}(d) where a Fermi surface process is shown. All  diagrams belonging to the Migdal{'s} approximation exhibit this behavior. The FSR Migdal-Eliashberg Eqs.\,(\ref{zFs}-\ref{deltaFs}) correspond to diagrams in the first row of Figs.\,\ref{feynman2} where additionally all processes take place at the Fermi surface as depicted in Figs.\,\ref{feynman2}(d-f). When $\alpha$ increases, processes with at least one line connecting states away from the ellipses in Figs.\,\ref{feynman2}(d-f) become increasingly relevant due to the enlarged phase-space, thus the AFB Eliashberg Eqs.\,(\ref{zAd}-\ref{phiAd}) become relevant. 

On the other hand, for crossing diagrams, like e.g. Fig.\,\ref{feynman2}(g) and (h), particles scatter to different intermediate states. This means that for a Fermi surface without good nesting or when the {electron-phonon} interaction is not peaked at small-$q$, some of the intermediate states need to lie away from the Fermi level. When $\alpha$ is small, the limited available phase-space for such processes, as can be inferred by Figs.\,\ref{feynman2}(i-j), results in the suppression of the crossing diagrams. If we additionally consider only Fermi-surface processes, this suppression is further enhanced and the contribution of these diagrams becomes negligible. This is Migdal's theorem. As already stated the theorem can break down for small-$q$ phonons or well-nested Fermi surfaces. We will not address cases where the electron-phonon interaction is momentum dependent here. Apart from potential changes in the scaling of the vertex correction \cite{Madhukar1977}, reduced dimensionality may generally also lead to deviations from Migdal's theorem due to enhanced nesting conditions. Therefore, in 1D where nesting is perfect Migdal's theorem should not hold \cite{Allen1983}.

For arbitrary values of $\alpha$ but assuming only near Fermi surface processes, one retrieves the momentum conservation constraint of Eq.\,(\ref{momcons}) and the corresponding non-adiabatic Eliashberg equations in this approximation are given by Eqs.\,(\ref{iso1}-\ref{iso2}). By construction and given the preceding discussion, these equations can have a significant effect only in the case of enhanced nesting conditions. On the contrary, the solution of the full vertex-corrected Eliashberg Eqs.(\ref{z}-\ref{phi}) presents the most general case where all processes are considered and no assumption is made on the value of $\alpha$. Thus, all diagrams are properly taken into account in this case.

We now turn to discuss our numerical solutions in the next Section.

\section{Results}\label{scResults}

The full parameter space introduced in Sec.\,\ref{scSimp} is too big for a complete numerical analysis, especially for 3D systems. We confine ourselves to a characteristic phonon frequency of $\Omega=50\,\mathrm{meV}$, fix the electronic bandwidth at $W=1.5\,$eV and impose an isotropic coupling, $f_{\mathbf{q}}=1_{\mathbf{q}}$. Due to the fact that existing theories of non-adiabatic superconductivity are confined to the Fermi surface\,\cite{Pietronero1992_2,Paci2001,Benedetti1994} we perform a variation in $\lambda$ to make a comparison easier. By means of Eq.\,(\ref{g0}) we have a tool to translate this variation into the expansion parameter $|g_0|^2$, which is more meaningful in our full-bandwidth treatment. For the here-presented results we are interested in {reaching a qualitative understanding rather than quantitative absolute numbers,} since we stay on a model basis.

\subsection{Dimensionality}\label{scDimension}

We want to compare our self-consistent full-bandwidth non-adiabatic results with {AFB}, adiabatic {FSR}  {\blue and non-adiabatic FSR isotropic} calculations. For this purpose we consider the maximum superconducting gap as function of coupling strength for different degrees of non-adiabaticity $\alpha$, repeated for each possible dimensionality. Further we fix the temperature at $T=20\,\mathrm{K}$ in the current section.

\subsubsection{3D systems}

\begin{figure}[h!]
	\centering
	\begin{overpic}[width = 1\columnwidth, clip, unit=1pt]{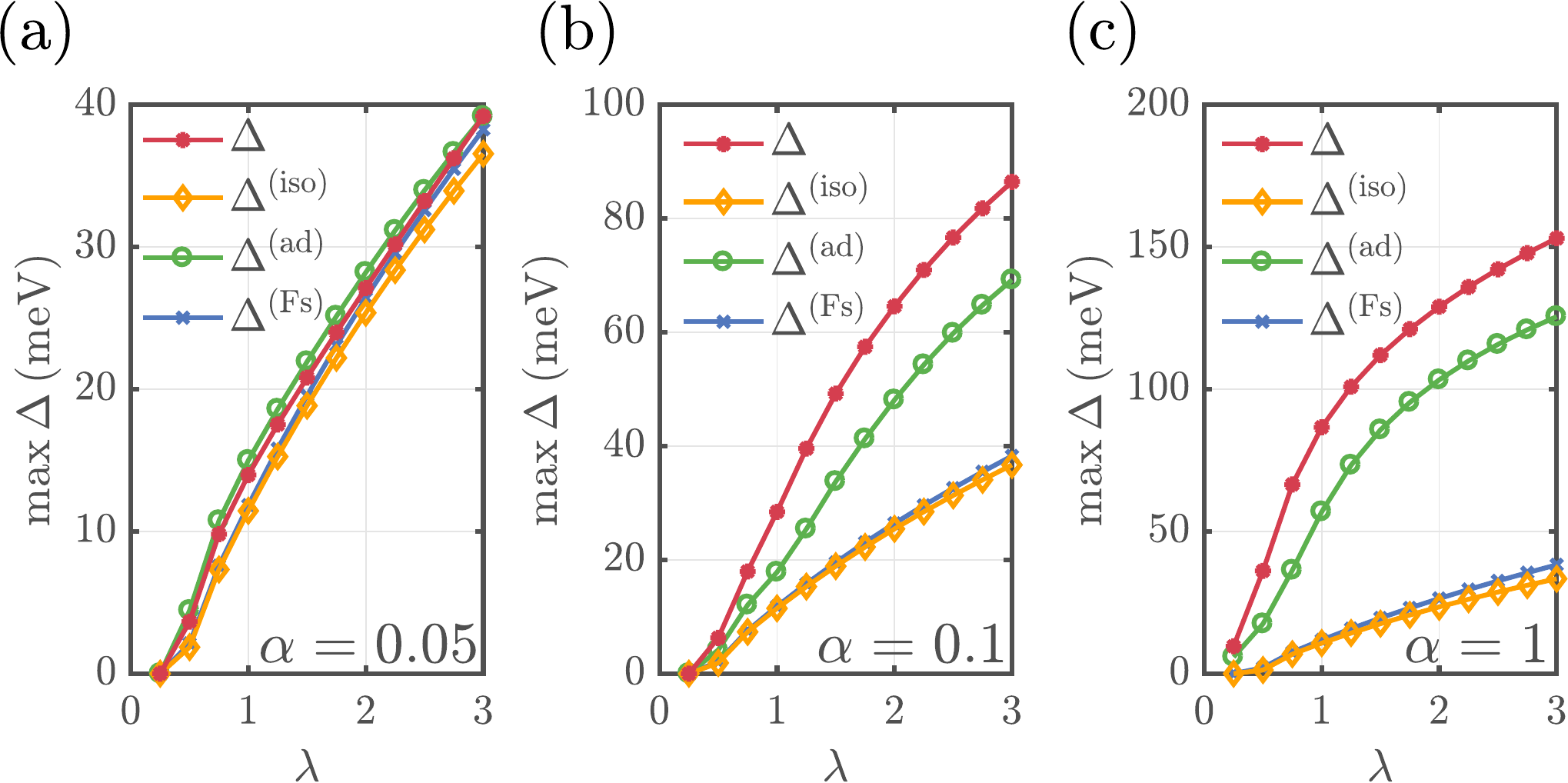}
	\end{overpic}
	\caption{Maximum superconducting gap as function of coupling strength $\lambda$ in 3D systems. Red stars, green circles, blue crosses {\blue and yellow diamonds} show results for our {\blue complete} non-adiabatic algorithm, the {AFB} calculation and {FSR} equations {\blue and the FSR isotropic approximation to the non-adiabatic equations}, respectively. {(a)} Non-adiabaticity parameter $\alpha=0.05$; {(b)} $\alpha=0.1$; and {(c)} $\alpha=1$.}
	\label{example3D}
\end{figure}

It is well established that Migdal's theorem is valid in three spatial dimensions, hence we expect vertex corrections to be small, provided that $\alpha \ll 1$. Choosing $\alpha=0.05$ and $c_x=c_y=c_z=1$, we test this aspect in Fig.\,\ref{example3D}{(a)} by comparing the adiabatic maximum gap (green circles) and the {FSR} results (blue crosses) with outcomes from our {\blue complete} non-adiabatic Eqs.\,(\ref{z}-\ref{phi}) (red stars) {\blue and the FSR isotropic non-adiabatic Eqs.\,(\ref{iso1}-\ref{iso2}) (yellow diamonds). Our 3D solutions for the latter approximation coincide almost perfectly with the ones obtained from FSR non-adiabatic calculations, as can be seen in all panels of Fig.\,\ref{example3D}. We will return to this point at the end of the section.}
As {is} directly apparent, the vertex corrections in such an adiabatic situation are indeed very small. Additionally we find good agreement also with $\Delta^{(\mathrm{Fs})}$, which points towards purely Fermi surface based Cooper pairing {and indicates that a FSR approximation is sufficient in this parameter regime}. The observations hold true for all couplings we tested, hence the essential ingredients are the 3D character of the system and smallness of $\alpha$. We checked that there is good agreement between all four approaches for $\alpha<0.05$.

Next we examine the effect of making the system less adiabatic, $\alpha=0.1$. In similar color code as before we show the outcomes in panel \ref{example3D}(b). For increasing coupling strength the difference between {FSR} and {AFB} calculations grows larger. Further we find for all $\lambda$ values that $\Delta^{(\mathrm{ad})}$ in this setup is an underestimation of the vertex-corrected maximum gap size $\Delta$. We observe that these trends persist when further increasing $\alpha$ in Fig.\ \ref{example3D}(c). Here $\Delta^{(\mathrm{Fs})}$ is far too small compared to non-adiabatic results. 
Note, that this property stems from the fact that the {FSR} theory of Eqs.\,(\ref{zFs}) and (\ref{deltaFs}) does not explicitly depend on $\xi_{\mathbf{k}}$, {\blue  since we here assume a momentum independent electron-phonon coupling $g_0$. Therefore it is independent of $\alpha$. In other words, due to taking $g_{\bf q}=g_0$, Eqs.\,(\ref{zFs}) and (\ref{deltaFs}) assume the form of the usual isotropic Eliashberg equations \cite{carbotte1990}.}
 Further, we find $\Delta>\Delta^{(\mathrm{ad})}$ throughout the whole range of couplings, while their relative difference stays comparatively small. In both panels {(b)} and {(c)} we get a deviation of $|(\Delta-\Delta^{(\mathrm{ad})})/\Delta|\sim20\%$ in the large coupling limit. This can be explained in terms of the expansion parameter $|g_0|^2$, which is smaller for panel {(c)} than for panel {(b)}. Hence there is an increased importance of non-adiabatic effects which, however, are weighted less. It is worth mentioning that Fig.\ \ref{example3D}(c) may be relevant to the situation encountered in H$_3$S where $\alpha\approx 1$ and $\lambda\approx 2$ \cite{Errea2015,Jarlborg2016} and in  SrTiO3$_3$ where $\alpha\gg 1$ and $\lambda\approx 0.1-0.4$ \cite{Marel2011}. 
 
{\blue {Previously,}  solving the linearized Eliashberg equations {for H$_3$S gave}  that $T_c$ is reduced when considering the full energy dependence of the electronic  DOS in comparison to using the constant DOS approximation \cite{Sano2016}.   These two approaches compare to the linearized versions of the here discussed AFB and FSR approaches. If we associate the superconducting gap maximum with the expected magnitude of the corresponding $T_c$, the results of Ref.\,\cite{Sano2016} regarding these two approaches are opposite to our findings shown in Fig.\,\ref{example3D}(c). In our case, we find that the AFB approach yields a higher $T_c$ as compared to FSR calculations for 3D systems. A possible reason for this difference may 
be the fact that the calculations shown in Fig.\,4 and Fig.\,5 of Ref.\,\cite{Sano2016} are one-shot, i.e.\ not self-consistent. Nevertheless, our findings are in line with results obtained for the case of di-hydrogen sulphide, H$_2$S \cite{Sano2016}.  In addition, vertex corrections were shown to reduce $T_c$ in these materials \cite{Sano2016}. This again is in contrast to our results shown in Fig.\ \ref{example3D}(c). 
We believe that this discrepancy is a consequence of the approximations adopted in Ref.\,\cite{Sano2016} when including the vertex corrections, especially taking the static limit of the vertex function. These approximations are similar to those introduced in Ref.\,\cite{Grimaldi1995a} which we discuss in more detail further below in Sec.\,\ref{scComparison}.}

\subsubsection{2D systems} 
\begin{figure}[h!]
	\centering
	\begin{overpic}[width = 0.975\columnwidth, clip, unit=1pt]{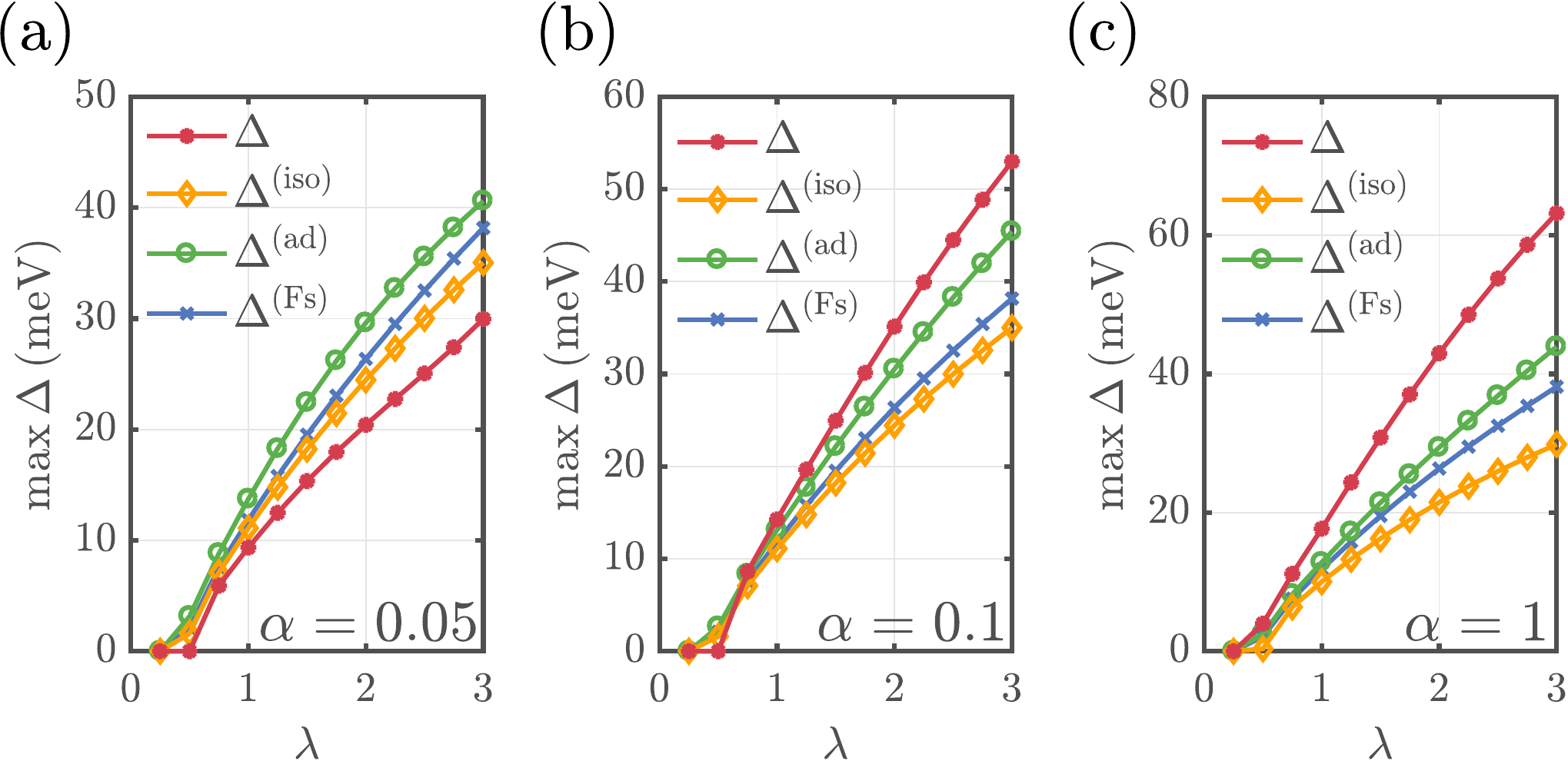}
	\end{overpic}
	\caption{Comparison of maximum superconducting gaps in 2D systems. Our results are self-consistently {computed} from the {\blue four approaches}	 discussed in this work, using {the same}	color code and degrees of non-adiabaticity $\alpha$ as in Fig.\,\ref{example3D}.}
	\label{example2D}
\end{figure}

Next we treat the two dimensional case by setting $c_x=c_y=1$ and $c_z=0$. For this situation there is no proof that vertex corrections are negligible, even for $\alpha \ll 1$. It is further{more} exceptionally interesting to examine possible effects in this setup, because many high-$T_c$ superconductors {are either quasi-2D \cite{Damascelli2003,Johnston2010} or pure 2D systems \cite{Logvenov2009,Qing-Yan2012,Lee2014,Yu2019}}. We choose similar parameters as before, $\alpha=0.05$, $\alpha=0.1$ and $\alpha=1$, the results of which we plot in panels {(a)}, {(b)}, and {(c)} of Fig.\,\ref{example2D}, respectively. The color code and choice of axes (coupling versus maximum superconducting gap) is the same as in  Fig.\,\ref{example3D}.
A common observation in all three panels of Fig.\,\ref{example2D} is a rather small deviation between results from {FSR} and {AFB} calculations. This hints towards small tendencies of {Cooper pairing away from the Fermi level} in the parameter space we examine here \cite{Aperis2018,Gastiasoro2019}. Turning to the maximum superconducting gap from our non-adiabatic theory we find $\Delta<\Delta^{(\mathrm{Fs})}<\Delta^{(\mathrm{ad})}$ for all couplings considered in Fig.\,\ref{example2D}{(a)}. By neglecting vertex corrections one therefore overestimates the gap size to an extent that depends on $\lambda$. As we discuss in Sec.\,\ref{scTrends} below, this overestimation is not restricted to the superconducting gap, but translates also to the transition temperature.

Before discussing the intermediate case of Fig.\,\ref{example2D}{(b)}, let us first turn to $\alpha=1$ in panel {(c)}. Here we find an opposite trend to before, i.e.\ an underestimation of the {non-adiabatic} gap {by}  both adiabatic algorithms {and} for all $\lambda$. Again referring to Sec.\,\ref{scTrends} {below,} this trend is similarly true for $T_c$. 
{An example {of a} superconductor where Fig.\,\ref{example2D}{(c)} may be relevant is FeSe/SrTiO$_3$ where $\alpha\approx 1.6-2$ and $\lambda\approx 0.2-0.4$ \cite{Lee2014,Rademaker2016,Aperis2018}. Despite the fact that we here assumed a plain electron-phonon coupling, instead of the small-$q$ interaction that is at play in FeSe/SrTiO$_3$, our results provide further support that the latter mechanism is capable of mediating the observed high-$T_c$. Moreover,}  our findings coincide qualitatively  with non-adiabatic, but {FSR} calculations carried out on small-$q$ scattering  {\blue (see also Sec.\,\ref{scComparison})} in C$_{60}$ compounds\,\cite{Pietronero1992_1,Pietronero1995}. 

Let us now turn to the intermediate situation $\alpha=0.1$ in Fig.\,\ref{example2D}{(b)}. For sufficiently large couplings we retrieve the situation of $\alpha=1$, while {results for} small $\lambda$ resemble more closely the situation in panel {(a)}. It can hence be argued that the transition from the adiabatic ($\alpha \ll 1$) to the non-adiabatic regime ($\alpha\sim1$) is smooth, where in an intermediate situation as in panel {(b)}, the over- or underestimation of $\Delta^{(\mathrm{ad})}$ and $\Delta^{(\mathrm{Fs})}$ with respect to vertex corrected results depends on the coupling strength only. {It is interesting to note, {lastly}, that cuprates \cite{Damascelli2003} and iron pnictides \cite{Johnston2010} {fall} in this intermediate regime.}

\subsubsection{1D systems} 

\begin{figure}[h!]
	\centering
	\begin{overpic}[width = 1\columnwidth, clip, unit=1pt]{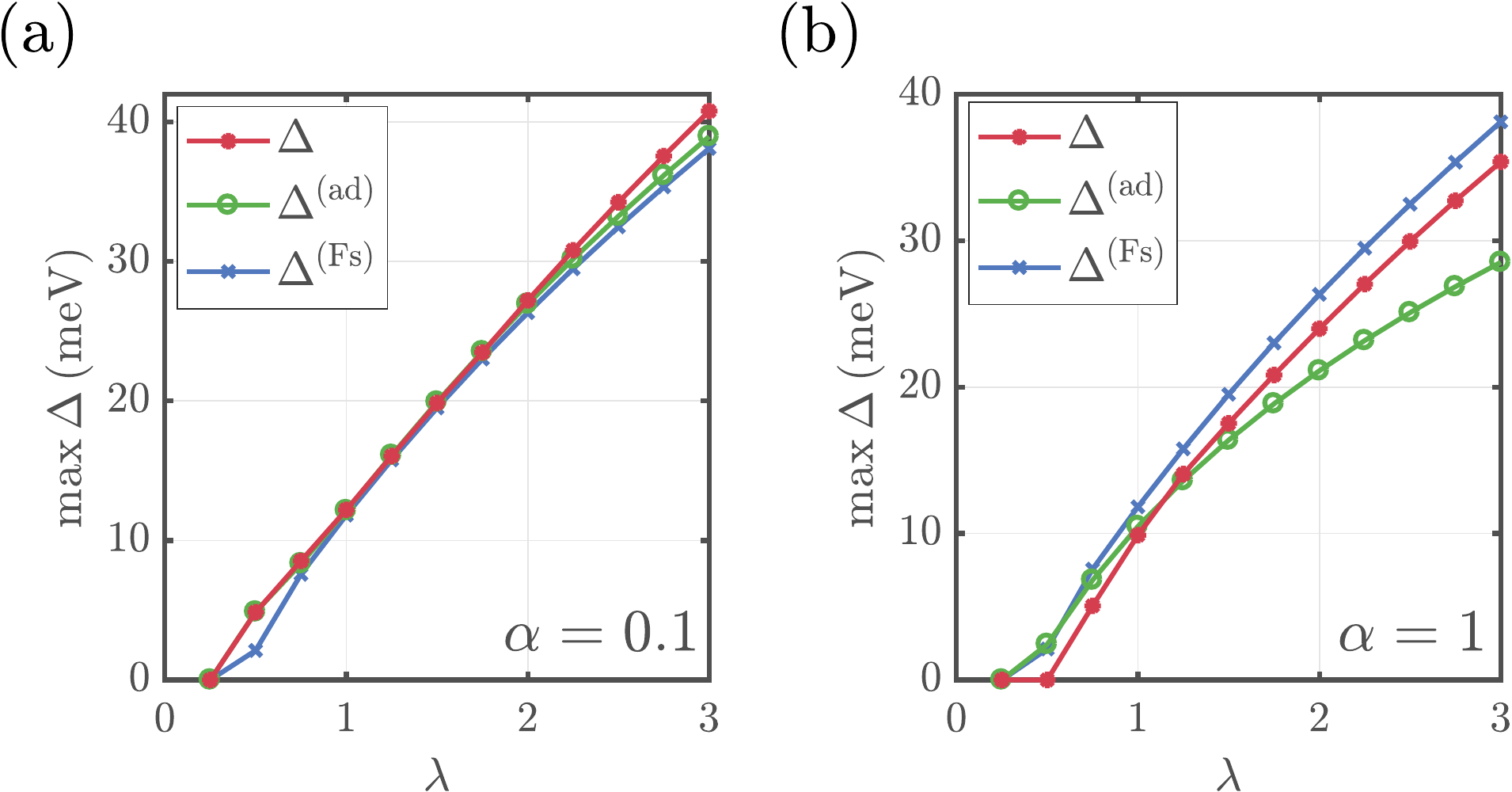}
	\end{overpic}
	\caption{Maximum superconducting gaps in 1D systems for $\Delta$, $\Delta^{(\mathrm{ad})}$ and $\Delta^{(\mathrm{Fs})}$ self-consistently {computed} from Eqs.\,(\ref{z}-\ref{phi}), (\ref{zAd}-\ref{phiAd}), and (\ref{zFs}-\ref{deltaFs}), respectively. The {same} color code {as in} Figs.\,\ref{example3D} and \ref{example2D} is used. {(a)} $\alpha=0.1$, {(b)} $\alpha=1$.}
	\label{example1D}
\end{figure}

{For completeness, we briefly discuss here results}
when lowering the system's dimensionality to 1D via $c_x=1$ and $c_y=c_z=0$. 
{\blue In 1D, it was not possible to solve the FSR isotropic approximation to the non-adiabatic Eliashberg equations (\ref{iso1}) and (\ref{iso2}) due to the phase-space constraint imposed by the prefactor $P$ of Eq.\,(\ref{momcons}).} In Fig.\,\ref{example1D} we show again the comparison of results from Eqs.\,(\ref{z}-\ref{phi}) with outcomes of Eqs.\,(\ref{zAd}-\ref{phiAd}) and Eqs.\,(\ref{zFs}-\ref{deltaFs}) {for 1D systems}, just as is done in Figs.\,\ref{example3D} and \ref{example2D}. 
{For $\alpha=0.1$ (the case $\alpha=0.05$ has similar trends), we observe very small effects due to vertex corrections [see Fig.\,\ref{example1D}{(a)}].}
Although a slight deviation from $\Delta$ can be found for large $\lambda$, all three curves lie almost on top of each other. From this we can conclude that {AFB}, and even {FSR} calculations in 1D are very accurate with respect to the maximum superconducting gap, provided that $\alpha\lesssim\mathcal{O}(0.1)$. Further we note that $\Delta>\Delta^{(\mathrm{ad})}>\Delta^{(\mathrm{Fs})}$ for all couplings shown.
An increased influence of non-adiabatic effects is found with enhanced $\alpha=1$ in panel \ref{example1D}{(b)}. When comparing $\Delta$ and $\Delta^{(\mathrm{ad})}$ the situation is similar as in Fig.\,\ref{example2D}{(b)}, i.e.\ the adiabatic results are an over- or an underestimation depending on the coupling strength. In addition we find a result unique to this 1D simulation: {Among our three approaches, FSR} calculations lead to the largest gaps {for all coupling strengths we consider}. 

{From Fig.\,\ref{example1D}{(a)} and Fig.\,\ref{example3D}{(a)} one sees that in 1D, Migdal's approximation stays valid for larger values of the non-adiabaticity parameter as compared to the 3D case. This result is unexpected since in 1D Migdal's theorem should not hold \cite{Allen1983}.
It is worth mentioning that in 1D the nesting properties of the Fermi surface (that now consists of two points) enhance the tendency of the system towards the formation of a Charge Density Wave (CDW) or Peierls instability \cite{Gruener1988}. In principle, such tendency should be taken into account by including both superconducting and CDW orders on equal footing in the Eliashberg equations. However, {doing this} in the presence of vertex corrections is out of the scope of the present work.}

{\blue \subsubsection{Discussion of FSR isotropic non-adiabatic approach}} 

We close this section with a comment on the validity of the FSR isotropic approximation to the non-adiabatic Eliashberg equations, namely Eqs.\,(\ref{iso1}-\ref{iso2}). We have solved these equations for 3D and 2D cases. Similar to what we found in 3D, these solutions follow closely the FSR adiabatic results in the 2D case, too. According to the analysis of Sec.\,\ref{scDiagrams}, the reason for this behavior is that Eqs.\,(\ref{iso1}-\ref{iso2}) neglect non-Fermi surface contributions. In addition, the small deviations between the FSR adiabatic and the FSR isotropic non-adiabatic results in 2D can be understood as a manifestation of slightly enhanced nesting conditions when dimensionality is reduced from 3D to 2D.
Taking also into account the big discrepancy between the non-adiabatic FSR isotropic results and the solutions of the complete non-adiabatic Eqs.\,(\ref{z}-\ref{phi}) we can safely conclude that the FSR isotropic approximation severely underestimates the effect of the first vertex correction and is therefore not valid.

\subsection{Possible implications for 2D systems}\label{scTrends}

\begin{figure}[h!]
	\centering
	\begin{overpic}[width = 1\columnwidth, clip, unit=1pt]{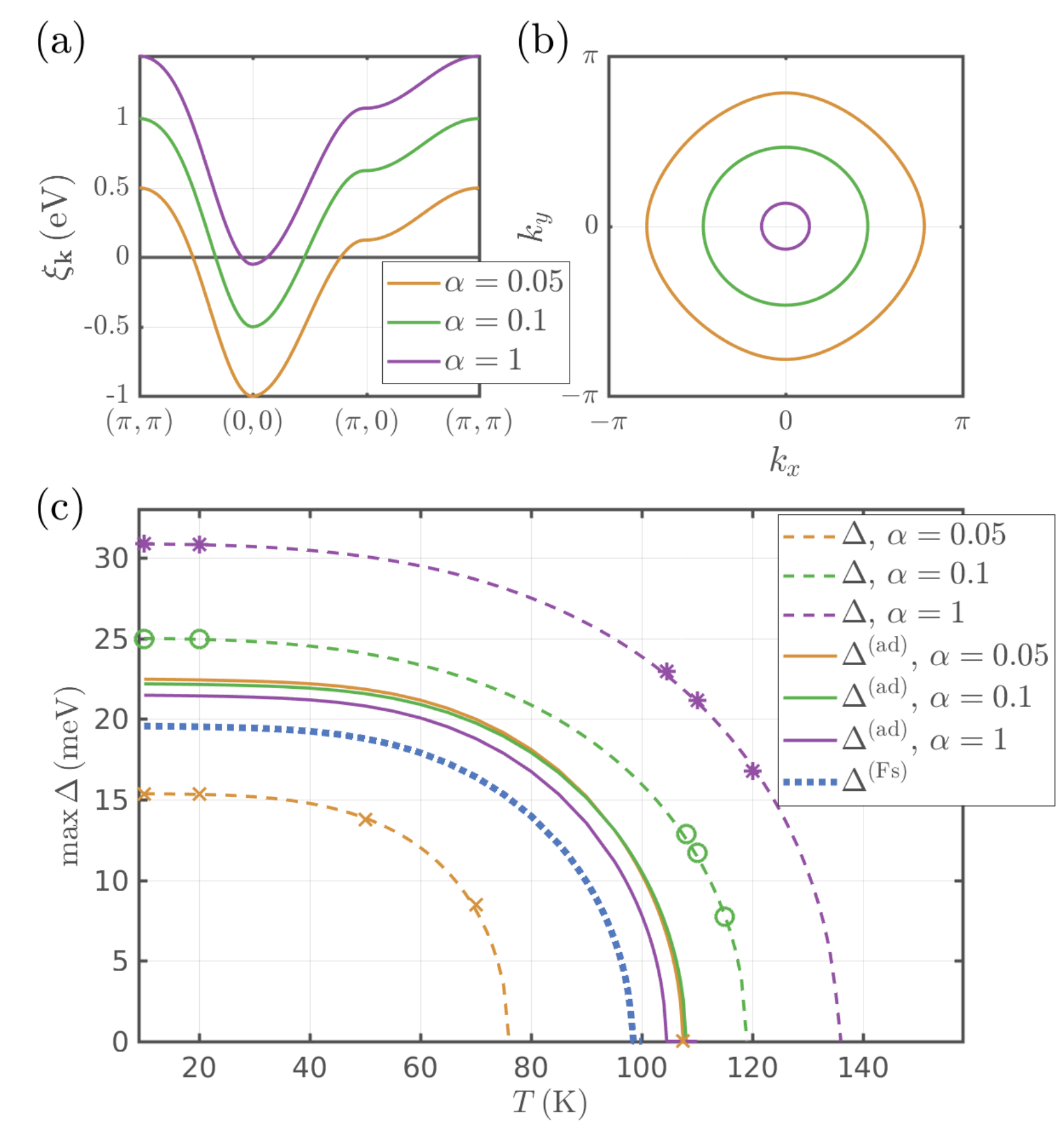}
	\end{overpic}
	\caption{{(a)} Electronic energies along high symmetry lines of the two dimensional BZ, shown for three different values of $\alpha$ as indicated in the legend.	{(b)} Fermi surfaces of the dispersions shown in panel {(a)}, using similar color code. {(c)} Self-consistently obtained superconducting gaps as function of temperature. Dashed and solid curves refer respectively to the non-adiabatic and the adiabatic {approach}. Our simulation results for the former are marked explicitly by crosses/circles/stars, while the dashed lines are obtained by a fit. Colors are the same as in panels {(a)} and {(b)}. For comparison we show outcomes of the {FSR} calculation by the dotted blue curve.}
	\label{gapclosing}
\end{figure}

From results in Sec.\,\ref{scDimension} we learn that, depending on system specifics such as dimensionality, an increase or decrease of the gap size due to non-adiabatic vertex corrections is possible. So far we did, however, {focus only on}
$\mathrm{max}\,\Delta$ and similar conclusions about the transition temperature $T_c$ cannot be drawn without further investigation. Strictly speaking, an enhancement in $\Delta$ might either lead to a larger $T_c$, or simply to a change in $2\Delta_0/k_BT_c$, with $\Delta_0=\underset{T\rightarrow0}{\mathrm{lim}}\,\big(\mathrm{max}\,\Delta\big)$. {T}his ratio is a common measure for how strongly coupled a superconductor is. Henceforth we closer examine 2D systems ($c_x=c_y=1$, $c_z=0$) with degrees of non-adiabaticity $\alpha=0.05$, $\alpha=0.1$ and $\alpha=1$, all at a coupling strength $\lambda=1.5$. Note that an exploration of $T_c$ in a comparable parameter space as in Sec.\,\ref{scDimension} is hardly feasible due to tremendous computational costs. Figure \ref{gapclosing}{(a)} shows the electronic energies along high-symmetry lines of the BZ with colors as indicated in the corresponding legend. The associated Fermi surfaces are drawn in panel {(b)} of the same figure. 
As evident from these graphs, by changing the chemical potential $\mu$ we accomplish shallow Fermi surface pockets, which result in an increased $\alpha$ since $\Omega=50\,\mathrm{meV}$ is kept constant. In Fig.\,\ref{gapclosing}{(c)} we draw a comparison of the temperature dependent maximum gap for different $\alpha$ and for each of {\blue the full bandwidth non-adiabatic, AFB and FSR} approaches as indicated in the legend. {\blue We do not include the FSR isotropic approximation to the non-adiabatic equations in this Section (see discussion at the end of Sec.\,\ref{scDimension}.}

Let us start with the results from {AFB} calculations, shown as solid lines with similar color code as in panels {(a)} and {(b)}. Our results for $T_c$ and $\Delta^{(\mathrm{ad})}(T)$ do not change significantly as function of $\alpha$. With growing non-adiabaticity we detect a small decrease in the transition temperature and {the zero-temperature superconducting gap } $\Delta^{(\mathrm{ad})}_0\simeq\Delta^{(\mathrm{ad})}(T=10\,\mathrm{K})$. When considering $\Delta^{(\mathrm{Fs})}$, shown as dotted blue line, the gap size and transition temperature decrease further, but are still in the same range as the full bandwidth calculations. The here-detected difference is due to Cooper pairing away from the Fermi level {\cite{Aperis2018,Gastiasoro2019}}, which is neglected when solving Eqs.\,(\ref{zFs}-\ref{deltaFs}), and is almost negligible in the current model system. From the discussion of our adiabatic calculations one can conclude that $T_c^{(\mathrm{ad})}\sim T_c^{(\mathrm{Fs})}\sim100$\,K and $\Delta^{(\mathrm{ad})}_0\sim\Delta^{(\mathrm{Fs})}_0\sim20\,$meV almost independently of $\alpha$, which then leads to $2\Delta^{(\mathrm{ad})}_0/k_BT^{(\mathrm{ad})}_c\sim2\Delta^{(\mathrm{Fs})}_0/k_BT^{(\mathrm{Fs})}_c\sim4.64$. This value reflects the strong coupling nature of {the system, within our chosen parameters.}

Inclusion of vertex corrections introduces some rather drastic changes compared to the adiabatic picture; the results of including these are shown as dashed lines in Fig.\,\ref{gapclosing}{(c)}. 
{For} $\alpha=0.05$ {(plotted in orange)}, both $\Delta_0\sim15\,$meV and $T_c\sim75\,$K are heavily reduced compared to the adiabatic counterparts, while $2\Delta_0/k_BT_c$ is enhanced to $\sim4.97$, giving rise to a {seemingly} more strongly coupled behavior. From Fig.\,\ref{example2D} we can readily see that $\Delta_0>\Delta_0^{(\mathrm{ad})}>\Delta_0^{(\mathrm{Fs})}$ when $\alpha=0.1$. In addition we see from data {plotted by the dashed green line} that $T_c$ is enhanced to approximately $120\,\mathrm{K}$, which gives again a stronger coupling $2\Delta_0/k_BT_c\sim4.84$ when combined with $\Delta_0\sim25\,$meV. For even shallower bands, $\alpha=1$, we find the dashed purple line in Fig.\,\ref{gapclosing}{(c)} to give $\Delta_0\sim31\,$meV and $T_c\sim135\,$K, and hence a very strong coupling situation of $2\Delta_0/k_BT_c\sim5.33$. If we consider changes in the transition temperature with the degree of non-adiabaticity for the vertex-corrected {theory}, our results for the current model system suggest that $T_c$ increases with $\alpha$. Interestingly, the trend for the maximum gap size in the limit $T\rightarrow0$ is reversed when comparing with the {AFB} outcomes, i.e.\ we find $\Delta_0^{(\mathrm{ad})}|_{\alpha=0.05}>\Delta_0^{(\mathrm{ad})}|_{\alpha=0.1}>\Delta_0^{(\mathrm{ad})}|_{\alpha=1}$ without and $\Delta_0|_{\alpha=0.05}<\Delta_0|_{\alpha=0.1}<\Delta_0|_{\alpha=1}$ with vertex corrections. We similarly find the largest (smallest) $T_c$ ($T_c^{(\mathrm{ad})}$) for the highest (lowest) $\alpha$.

It is interesting to see how the just-discussed trends in critical temperature compare to effects in the {effective}
electron-phonon coupling constant, which is given by 
\begin{align}
&\lambda_Z^{(\mathrm{ad}/\mathrm{Fs})} = \langle Z^{(\mathrm{ad/\mathrm{Fs}})}_{\mathbf{k},m=0} \rangle_{\mathbf{k}_F} |_{T>T_c}^{ } - 1 \label{adiabaticCoupling} ~.
\end{align}
This quantity is a measure for the coupling strength, renormalized due to interactions, and it is in general not equivalent to the {here-used} bare coupling $\lambda=1.5$, except for purely Fermi-surface based calculations. Note that a large number of Matsubara frequencies is needed to numerically confirm $\lambda_Z^{(\mathrm{Fs})}=\lambda$. In case of the non-adiabatic treatment we know from Eq.\,(\ref{z}) that contributions due to both scattering diagrams enter the mass renormalization in an additive way. It is therefore convenient to define
\begin{align}
\lambda_Z &= \langle Z_{\mathbf{k},m=0} \rangle_{\mathbf{k}_F} |_{T>T_c,\,T>T_c^{(\mathrm{ad})},\,T>T_c^{(\mathrm{Fs})^{ } }} - 1 \nonumber\\
&\equiv \lambda_Z^{(1)} + \lambda_Z^{(2)} \label{nonadiabaticCoupling} ~,
\end{align}
where $\lambda_Z^{(1)}$ and $\lambda_Z^{(2)}$ arise from the first and second diagram {in Fig.\,1}, respectively. For comparison it is useful to choose a temperature larger than $T_c^{(\mathrm{Fs})}$, $T_c^{(\mathrm{ad})}$ and $T_c$ when evaluating Eq.\,(\ref{nonadiabaticCoupling}). For full-bandwidth calculations we show results {obtained} for the coupling constant as function of $\alpha$ in Table\,\ref{table}.
\renewcommand{\arraystretch}{1.4}
\begin{table}[h!]
	\centering
	\caption{Calculated electron-phonon coupling constants as function of non-adiabaticity $\alpha$.
	The rows correspond to the {theoretical approach} used for calculating the results, {i.e.}\ either adiabatic or non-adiabatic full-bandwidth simulations, compare Eqs.\,(\ref{adiabaticCoupling}) and (\ref{nonadiabaticCoupling}).}\label{table}
	\begin{ruledtabular}
	\begin{tabular}{cccc } 
		& $\alpha=0.05$ & $\alpha=0.1$ & $\alpha=1$  \\ 
		\hline
		$\lambda^{(\mathrm{ad})}_Z$ & $1.4016$ & $1.5971$ & $1.4723$   \\
		\hline
		$\lambda^{(1)}_Z$ & $1.4015$ & $1.5972$ & $1.4671$   \\
		\hline
		$\lambda^{(2)}_Z$ & $-0.5496$ & $-0.0462$ & $0.4060$   \\
		\hline
		$\lambda_Z$ & $0.8519$ & $1.5510$ & $1.8731$   \\
	\end{tabular}
	\end{ruledtabular}
\end{table} 

From the first row we learn that non-Fermi surface {processes} give rise to coupling constants that are clearly different from the initial $\lambda=1.5$. 
{The trend of how $\lambda$ is modified as we increase the non-adiabaticity is not apparent.}
 We note, however, that $\alpha=1$ leads to the smallest deviations in both $|\lambda-\lambda^{(\mathrm{ad})}_Z|=|\lambda_Z^{(\mathrm{Fs})}-\lambda^{(\mathrm{ad})}_Z|$ and $|T_c^{(\mathrm{Fs})}-T_c^{(\mathrm{ad})}|$, compare Fig.\,\ref{gapclosing}{(c)}. Let us now consider solutions to the non-adiabatic Eqs.\,(\ref{z}-\ref{phi}). As already mentioned, $\lambda_Z^{(1)}$ corresponds to the first order scattering diagram of Fig.\,\ref{feynman}{(a)}, and hence can be compared to $\lambda_Z^{(\mathrm{ad})}$. 
 By comparing the first and second row in Table\,\ref{table} we observe that {indeed} $\lambda_Z^{(\mathrm{ad})}\simeq\lambda_Z^{(1)}$. {However, the deviation between these two quantities grows as $\alpha$ increases. This is due to the increasing importance of the feedback of the non-adiabatic terms on the adiabatic ones within the self-consistent iterative loop.} 
 The vertex corrections introduce the correction $\lambda_Z^{(2)}$, which according to the third row in Table\,\ref{table} increases significantly with $\alpha$. For the most adiabatic situation shown here, $\alpha=0.05$, we find $\lambda_Z^{(2)}<0$, hence the overall coupling constant is drastically reduced. In case of $\alpha=0.1$ we similarly get a negative $\lambda_Z^{(2)}$, but the magnitude is very small. Therefore the sum $\lambda_Z=\lambda_Z^{(1)}+\lambda_Z^{(2)}$ is closer to the bare coupling constant $\lambda$. Increasing the non-adiabaticity further to $\alpha=1$, the contributions due to the second order diagram become non-negligibly positive and give rise, together with $\lambda_Z^{(1)}$, to a  stronger electron-phonon coupling constant.

{Through} a closer inspection of the results for $\lambda_Z$ we can give a qualitative explanation of trends for the critical temperatures as they are observed in Fig.\,\ref{gapclosing}{(c)}. In the {AFB} calculations all couplings $\lambda_Z^{(\mathrm{ad})}$ lie within a narrow window of $\pm0.1$ around the bare coupling $\lambda=1.5$. It is therefore intuitive that corresponding values of $T_c^{(\mathrm{ad})}$ do also not differ drastically from each other. When solving the non-adiabatic  Eliashberg equations  (\ref{z}-\ref{phi}) we get an additional coupling contribution $\lambda_Z^{(2)}$, which heavily depends on $\alpha$. For the most adiabatic situation the effective  final coupling $\lambda_Z$ is comparatively weak, hence the critical temperature is smallest for $\alpha=0.05$. By increasing the degree of non-adiabaticity, we get {pairing contributions from $\lambda_Z^{(2)}$ which make the overall coupling stronger}. This, in turn, {leads to a rather pronounced enhancement} {of} $T_c$, as shown in Fig.\,\ref{gapclosing}{(c)}.\\

{\blue \subsection{Static \textit{vs.}\ dynamical limit and the momentum structure of the vertex function }\label{scComparison}}

As we discussed already, due to the fact that a direct numerical solution of the Eliashberg equations with the vertex corrected self-energy of Eq.\,(\ref{selfen}) has so far been missing,  {much} 
of the current understanding on the subject relies on a series of approximations. 
We have analyzed previous approaches on the simplification of the Eliashberg equations and their solution like e.g.,\ the FSR and isotropic approximations in the previous Sections. 
Another frequent approach, which we address here, is to introduce {further} assumptions in order to simplify the momentum and frequency structure of the vertex function itself. We will discuss this approach by focusing on two seminal papers by Pietronero \textit{et al}.\ \cite{Pietronero1995a,Grimaldi1995a}. 

It is worth pointing out that these works consider the normal state vertex function and neglect the self-consistent renormalization of the vertex due to the full electron propagator, e.g.\ no backreaction of the superconducting gap or the mass renormalization is taken into account. In other words, the Eliashberg equations in Ref.\,\cite{Grimaldi1995a} are derived from an electron self-energy that contains the diagram of Fig.\,\ref{feynman}(a) for the Migdal part and the diagram of Fig.\,\ref{feynman2}(g) for the vertex correction part. In contrast, we solve for the full self-energy of Fig.\,\ref{feynman}. 
In addition, despite the fact that the bare electron-phonon interaction in Ref.\,\cite{Grimaldi1995a} is taken as peaked at small-$q$, the derived Eliashberg equations correspond to isotropic, FSR Eliashberg equations, in our notation. This is due to the fact that the vertex function that contains the bare interaction, $g_q$, is averaged over Fermi surface momenta before it is plugged into the Eliashberg equations \cite{Grimaldi1995a,Pietronero1992_1,Pietronero1995}. Here, although we consider an isotropic bare interaction ($f_{\bf q}=1$ in Eq.\,(\ref{g0})), our numerical solutions do include the full momentum dependence of the resulting self-consistent vertex function. We leave the case of a momentum-dependent bare interaction for a future investigation.

By averaging the vertex function over both momentum and frequency, Pietronero \textit{et al}.\ found that the vertex function is positive if it peaks at small-$q$, whereas it is negative when the vertex function is weakly momentum dependent, i.e.\ it is almost isotropic \cite{Pietronero1995a}. Moreover, by approximating the vertex function as peaked at small-$q$, they studied the behavior of the vertex correction in the static and the dynamical limit. In the static limit, one first takes $\omega_m\rightarrow0$ and then ${\bf q}\rightarrow0$ in the vertex function, whereas in the dynamical limit one takes the limits in reversed order ${\bf q}\rightarrow0, \omega_m\rightarrow0$. They found that the static limit leads to a negative vertex function  \cite{Pietronero1995a} in agreement with prior results in the small-$q$ limit \cite{Grabowski1984}. In contrast, the vertex function was found positive in the dynamical limit \cite{Pietronero1995a}. 

\begin{figure}[h!]
	\centering
	\begin{overpic}[width = 0.98\columnwidth, clip, unit=1pt]{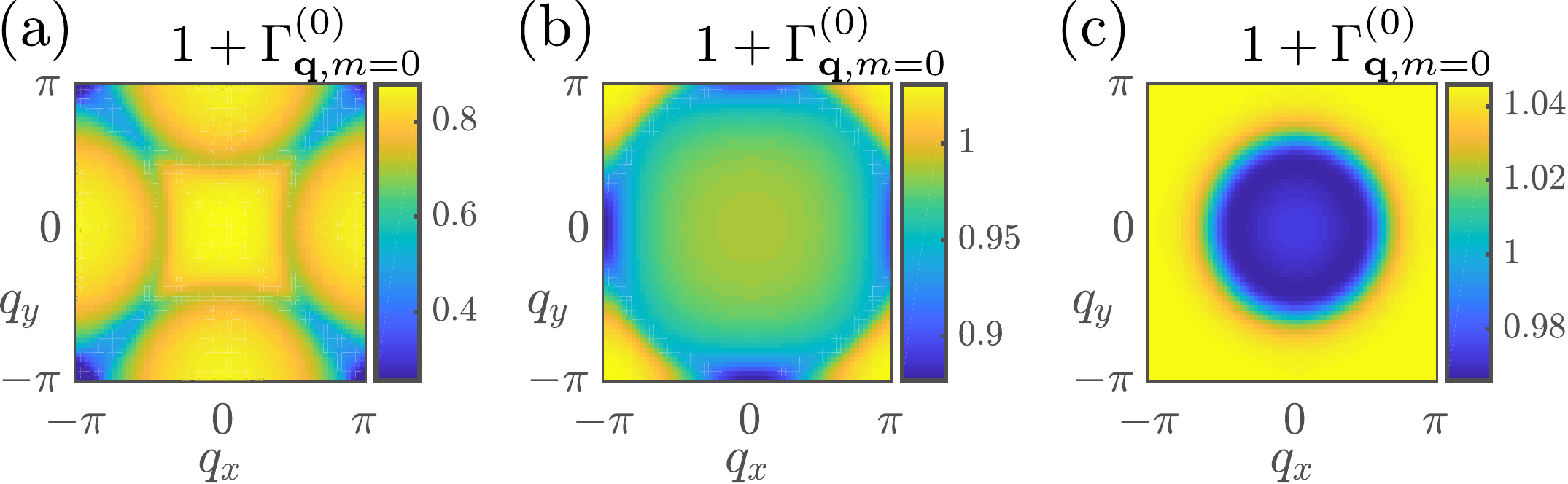}
	\end{overpic}
	\caption{Self-consistently calculated vertex renormalization function $1+\Gamma^{(0)}_{\mathbf{q},m=0}$ in the full {2-dimensional} BZ. {Panels} (a), (b), and (c) show results for $\alpha=0.05$, 0.1, and 1, and $\lambda=1.5$ corresponding to the 
	panels (a)--(c) of Fig.\,\ref{example2D}, respectively.}
	\label{fig7}
\end{figure}

Given that our own numerical solutions are free of any approximation to the vertex function, we turn to compare our results with these previous findings and discuss briefly the validity of such approximations.
For this purpose, we rewrite Eq.\,(\ref{selfen}) as,
\begin{align}
\hat{\Sigma}_{{\bf k}m} &= T\sum_{{\bf k}'m'}V_{m-m'}\hat{\rho}_3 \hat{G}_{{\bf k}',m'}\hat{\rho}_3 \Bigl( 1 +  \hat{\Gamma}_{\mathbf{k},\mathbf{k}',m,m'}\Bigl) 
\end{align} 
where the term in the parenthesis is the $2\times2$ matrix vertex renormalization function and $V_{\mathbf{k}-\mathbf{k}',m-m'}\equiv V_{m-m'}$ since in this work we have taken an isotropic bare interaction. After some straightforward algebra we find that the corresponding vertex renormalization function is given by $(-1)^\nu(1+\Gamma_{\mathbf{k},\mathbf{k}',m,m'})$, where $\nu=1,$ 0, {or} 0 for {the} $\phi$, $Z$, and $\chi$ channel, respectively. For the purely adiabatic part, $V_{m-m'}$, we set $m=m'$ and arrive at the equation for the vertex function:
\begin{align}\no
\Gamma^{(0)}_{\mathbf{q},m} &= -T \sum_{m''} V_{m'-m''} \sum_{{\bf k}''} \Bigl( \gamma_{\mathbf{k}'',m''}^{(Z)} \gamma_{\mathbf{k}''+\mathbf{q},m''}^{(Z)}\\\label{vertex0}
&~~~ - \gamma_{\mathbf{k}'',m''}^{(\chi)} \gamma_{\mathbf{k}''+\mathbf{q},m''}^{(\chi)}  + \gamma_{\mathbf{k}'',m''}^{(\phi)} \gamma_{\mathbf{k}''+\mathbf{q},m''}^{(\phi)}\Bigl) ,
\end{align}
where $\mathbf{q}=\mathbf{k}-\mathbf{k}'$ so that $\Gamma^{(0)}_{\mathbf{k},\mathbf{k}',m}\equiv \Gamma^{(0)}_{\mathbf{q},m}$.
Focusing on the 2D case and using our self-consistently calculated full Green's function, we find the momentum and frequency dependent vertex function. In order to investigate the resulting self-consistent momentum structure of the vertex correction,  we plot in Fig.\,\ref{fig7} the (${\bf q}$,$m$=0)-dependent vertex function of Eq.\,(\ref{vertex0}) for  $\alpha = 0.05$, $0.1$, and 1 that corresponds to panels Fig.\,\ref{example2D}(a-c) when $\lambda=1.5$. We note that, in this context, taking $w_m\rightarrow0$ in Eq.\,(\ref{vertex0}) does not correspond to taking the static or dynamical limit, since the vertex has already been calculated with full frequency and momentum dependence. 

As seen in Fig.\,\ref{fig7}, the resulting vertex function for the cases $\alpha =0.1$ and 1 where the vertex correction enhances $T_c$ exhibits little momentum anisotropy and by no means has a small-$q$ structure. In the case $\alpha = 0.05$ where the corrections suppress superconductivity as compared with the adiabatic case, the vertex is significantly momentum dependent but again it is not small-$q$. Given these results, we thus conclude that the vertex function cannot be \textit{a priori} approximated as small-$q$. However, one cannot exclude that this might be possible for a bare vertex that is already strongly peaked at small-$q$ \cite{Grimaldi1995a,Varelogiannis1998}, although this point deserves further investigation. 
 From Fig.\,\ref{fig7} one can also observe that as we increase the non-adiabatic ratio $\alpha$, the vertex function not only becomes overall more positive but also less momentum anisotropic. Hence, we see that the more isotropic the resulting vertex function is, the more positive its effect on the pairing becomes. Interestingly, these results are opposite to those of Ref.\,\cite{Pietronero1995a}.
  
Lastly, based on our results we cannot rule out the validity of neither the static nor the dynamical limit but we believe that the choice of either limit {depends}  on the specifics of the system under study. However, we point out that the vertex function $\Gamma^{(0)}_{\mathbf{q},m=0}$ appears to have a negative {effect} when $\alpha$ is very small, as can be seen clearly in Fig.\,\ref{example2D}(a)  for the 2D case. In combination with the fact that the FSR isotropic approximation to the non-adiabatic Eliashberg equations seems to systematically produce a negative correction (see e.g.,\ Figs.\,\ref{example3D} and \ref{example2D}) we are lead to believe that the static limit may be more relevant to systems where processes away from the Fermi level are not important, i.e. systems with low non-adiabaticity.

\section{Conclusion and outlook}\label{scConclusion}

We have investigated the influence of vertex corrections to the electronic self-energy on phonon-mediated,  anisotropic and full-bandwidth Eliashberg theory. To our knowledge the present investigation is the first numerical study to approach the challenge of non-adiabaticity in superconductors without involving further approximation, such as Fermi surface averages\,\cite{Pietronero1992_1,Pietronero1992_2,Botti2002,Boeri2002,Durajski2016} or momentum space clustering\,\cite{Hague2003,Hague2005}. 
Our calculations numerically confirm the validity of Migdal's theorem for 3D systems. For this dimensionality we have found, within the explored parameter space, that the {AFB} calculations always resemble the vertex corrected results to a good degree regardless of $\alpha$, which is however by no means true for {FSR} calculations. Contrarily, the observed trends in 2D systems suggest that non-adiabatic contributions are only negligible for rather small coupling strengths. We found that adiabatic results for the  {superconducting gap and $T_c$ can both be an over- or underestimation {of the vertex corrected ones}. In 1D systems we observe all corrections due to non-adiabaticity to be rather small. Adiabatic {FSR}, as well as {AFB} calculations start to become less accurate for $\alpha\gtrsim\mathcal{O}(1)$ and large couplings.} {\blue We have also investigated to what extend the full non-adiabatic equations can be approximated by simpler isotropic and FSR counterparts so as to reduce the huge computational effort. Our results prove that this is not possible and that the full non-adiabatic equations need to be solved, instead. 

Lastly, we have analyzed the momentum structure of the calculated vertex function and compared it with previous solutions that were obtained in the small-$q$, static or dynamical limit. Notably, our results indicate that in non-adiabatic systems vertex corrections can enhance superconductivity more efficiently when the resulting vertex function is less anisotropic, in contrast to previous findings \cite{Pietronero1995a}, {obtained under more restrictive assumptions}. Overall, our results emphasize the importance of going beyond previously employed approximations in estimating the impact of vertex corrections on the superconducting properties.}

{The model introduced in Sec.\,\ref{scSimp} gives rise to a huge parameter space, which we have partially explored here.} 
There is a large amount of extensions 
as well as applications, that go beyond the scope of {the present} work. {One example is to study anisotropy effects, i.e.\ cases when the bare pairing interaction is momentum dependent \cite{Aperis2018,Schrodi2018}. This includes electronic pairing mechanisms. For example, it has been shown that no analogue of Migdal's theorem exists for spin fluctuations, which means that vertex corrections are of similar order as the bare vertex\,\cite{Hertz1976}. Lastly, the here-presented methods can in principle be made compatible with {\em ab initio} input from Density Functional Theory (DFT) based calculations \cite{Aperis2015,Bekaert2019,UppSC}. The current caveat in this respect is of numerical nature, since the computational complexity (see Appendix \ref{appNum}) prohibits the usage of multiple electronic bands and very dense momentum grids. Nevertheless, such a non-adiabatic treatment would be an important step towards an even more realistic description of {superconductivity in actual} materials.}

\begin{acknowledgments}
	This work has been supported by the Swedish Research Council (VR), the R{\"o}ntgen-{\AA}ngstr{\"o}m Cluster, the K.\ and A.\ Wallenberg Foundation (grant No.\ 2015.0060) and the Swedish National Infrastructure for Computing (SNIC).	
\end{acknowledgments}

\bibliographystyle{apsrev4-1}
%

\appendix

\section{Numerical details}\label{appNum}

Here we describe some important aspects needed for the numerical implementation of Eqs.\,(\ref{z}-\ref{phi}).

\paragraph{Computational complexity.}
For discussing the costs of solving the multi-component, non-adiabatic, anisotropic and full-bandwidth Eliashberg equations let us denote the number of Matsubara frequencies by $N_{\mathcal{M}}$. Further let $N_x$, $N_y$, and $N_z$ be the number of $\mathbf{k}$-points along the three spatial directions. The most expensive part of the computation in Eqs.\,(\ref{z}-\ref{phi}) is by far the double summations over momenta and frequencies in the non-adiabatic terms, which correspond to diagram Fig.\,\ref{feynman}{(b)}. This is why we consider all remainders as constant, complexity-wise. Regarding the wave vectors, we need to execute one loop explicitly, inside of which we carry out a Fourier convolution, which gives roughly a scaling of $\mathcal{O}\left(N_x^2N_y^2N_z^2\log(N_xN_yN_z)\right)$. 

A similar scaling of $\mathcal{O}(N_{\mathcal{M}}^2\log(N_{\mathcal{M}}))$ can be achieved for the number of Matsubara frequencies by employing Fourier summation techniques. Putting these observations together we get the scaling
\begin{eqnarray}
\mathcal{O}\left( N_x^2N_y^2N_z^2N_{\mathcal{M}}^2\log(N_xN_yN_zN_{\mathcal{M}}) \right) , \label{compl}
\end{eqnarray}
for solving the non-adiabatic Eliashberg equations. 
This result is only correct in the sense of computational complexity, the actual computation is even heavier since several operations must be carried out multiple times. From expression (\ref{compl}) it is clear that all symmetries in momenta and frequencies must be exploited to be able to solve the non-adiabatic equations in a variational manner as is done in Sec.\,\ref{scResults}.

\paragraph{Tail fitting.}

In the iterative cycle of solving the equations a faithful interpolation along the frequency axis is needed. To be more specific, after each iteration we have access only to the current grids of functions $Z_{\mathbf{k},m}$, $\chi_{\mathbf{k},m}$ and $\phi_{\mathbf{k},m}$. In momentum space this is sufficient to solve for the same quantities in the next iteration, due to periodicity in the BZ. However, along the frequency axis we have a different situation, since we need in particular the term $\vec{\gamma}_{\mathbf{k}''-\mathbf{k}'+\mathbf{k},m''-m'+m}$, where each one of $m$, $m'$ and $m''$ can take values in $[-\mathcal{M},\, \mathcal{M}-1]$. Here we use $\mathcal{M}\in\mathbb{N}$ to denote the numerical cutoff for the Matsubara frequency grid. The value of $\mathcal{M}$ is found from convergence studies in the number of Matsubara frequencies. There is no periodicity in $\omega_m$, so the grid has to be enlarged to $[-3\mathcal{M},\,3\mathcal{M}-1]$ by performing a reliable extrapolation. One possible way of achieving this is to fit the tails on the frequency axis to a polynomial in $1/|\omega_m|$. This procedure works in a reliable way, provided that the tails of functions in Eqs.\,(\ref{z}-\ref{phi}) are sufficiently decayed already on the interval $[-\mathcal{M}, \, \mathcal{M}-1]$.

\end{document}